\documentclass[aps,prd,amsmath,amssymb,showpacs]{revtex4}
\usepackage{bm}

\def\journal#1#2#3#4{{#1} {\bf #2}, #3 (#4)}
\newcommand{\be}{\begin{equation}}
\newcommand{\ee}{\end{equation}}
\newcommand{\bea}{\begin{eqnarray}}
\newcommand{\eea}{\end{eqnarray}}
\newcommand{\hf}{\frac12}
\newcommand{\nn}{\nonumber\\}
\def\eq#1{(\ref{#1})}
\def\la{\langle}
\def\ra{\rangle}
\def\Tr{{\mathrm{Tr}}}

\def\mr#1{{\mathrm{#1}}}
\def\v#1{{\bm{#1}}}
\def\br{\hskip-6pt/}
\def\bre{\hskip-4pt/}
\def\fd#1#2{\frac{\delta#1}{\delta#2}}
\def\fdd#1#2#3{\frac{\delta^2#1}{\delta#2\delta#3}}
\def\ab{\bar a}

\def\at{\bar T}

\def\ha{{\hat a}}

\def\hhi{{\hat\phi}}
\def\hj{{\hat j}}

\def\hs{\hat\sigma}
\def\hsi{\hat\psi}
\def\hsib{\hat{\bar\psi}}
\def\htG{{\hat{\tilde G}}}
\def\htGi{{\hat{\tilde G}}^{-1}}
\def\hta{{\hat\eta}}
\def\htab{{\hat{\bar\eta}}}
\def\hA{{\hat A}}
\def\hD{{\hat D}}
\def\hDD{{\hat\Delta}}
\def\hG{{\hat G}}

\def\hJ{{\hat J}}
\def\hW{{\hat W}}
\def\etab{{\bar\eta}}
\def\jb{\bar j}

\def\Je{J^\mr{ext}}
\def\psib{\bar\psi}

\def\tG{{\tilde G}}

\begin{document}
\title{Quantum-classical crossover in electrodynamics}
\author{Janos Polonyi}\email{polonyi@fresnel.u-strasbg.fr}
\affiliation{Theoretical Physics Laboratory, Louis Pasteur University, Strasbourg, France}
\date{\today}
\begin{abstract}
A classical field theory is proposed for the electric current and the electromagnetic
field interpolating between microscopic and macroscopic domains.
It represents a generalization of the density functional for the dynamics of the
current and the electromagnetic field in the quantum side of the crossover
and reproduces standard classical electrodynamics on the other side.
The effective action derived in the closed time path formalism and the
equations of motion follow from the variational principle.
The polarization of the Dirac-see can be taken into account in the quadratic 
approximation of the action by the introduction of the deplacement field
strengths as in conventional classical electrodynamics. Decoherence appears
naturally as a simple one-loop effect in this formalism. It is argued
that the radiation time arrow is generated from the quantum boundary conditions
in time by decoherence at the quantum-classical crossover and the Abraham-Lorentz force
arises from the accelerating charge or from other charges in the macroscopic or
the microscopic side, respectively. The functional form of quantum renormalization 
group, the generalization of the renormalization group method for the density 
matrix, is proposed to follow the scale dependence through the 
quantum-classical crossover in a systematical manner.
\end{abstract}

\maketitle

\section{Introduction}
Classical systems traditionally serve as starting points for quantization.
But the opposite order, namely the construction of macroscopic physics from quantum 
principle, is needed to understand the great division line in Physics, the
quantum-classical crossover. One expects no surprise from the
traditional way of thinking because the usual equations of motions are supposed to
be recovered in the narrow wave-packet limits according to Ehrenfest theorem. 
Thought formally correct, this theorem does not guarantee that the original,
naive scenario of classical particles or bodies is recovered because the 
relativistic and many-body aspects of quantum physics introduce correlations 
in the narrow wave-packet limit which
render the picture of classically localized particles and bodies more involved.
The goal of this paper is to derive classical dynamics, namely action
and variational equations of motion for excitations in QED. 

The dynamics of expectation values can be obtained by means of the closed time path 
(CTP) formalism proposed by Schwinger long time ago \cite{schw}. One expects
that the dynamics for the electric current and the electromagnetic field 
will be governed by the action
\be\label{ced}
S=-\sum_iM_i\int_{x_i}ds-\sum_ie_i\int_{x_i}A_\mu(x)dx^\mu
-\frac{1}{4}\int d^4x(\partial_\mu A_\nu(x)-\partial_\nu A_\mu(x))^2+S_Q
\ee
involving the world lines $x_i(s)$ of charge $e_i$, mass $M_i$ and the
vector potential $A_\mu(x)$ and $S_Q$ standing for the corrections due to quantum fluctuations.
But there is a negative result conjecturing that no variational principle is available for 
the derivation of the equations of motion \cite{jordan} and in fact, up to our
knowledge no positive result had been communicated in this direction ever since. 
It will be shown below that the action \eq{ced} together with its
variational principle can actually be derived.
The term $S_Q$ in Eq. \eq{ced} represents the systematical improvement of
classical electrodynamics by taking into account quantum effects arising from
the polarization of the Dirac-see without interaction, vacuum polarizations
due to electromagnetic interactions and boundary conditions in time.

Once the action governing the expectation values is constructed a number 
of interesting questions open up. The first set of questions 
concerns the form of the action. How can we separate off the many-body 
aspects from a particle propagating in the non-interacting Dirac-see and what
corrections will be added to the free mechanical action, the fist term on the right hand side
of Eq. \eq{ced}? The dynamics of the expectation values of local operators should give a 
better insight into the quantum-classical transition regime because the space-time resolution 
of the expectation value of a local operator is limited by the UV cutoff only. How can we 
separate microscopic and macroscopic effects in the expectation values, or 
preferably in the action? The effects of the polarization of the Dirac-see are
included in the quantum corrections to the action. Does this happen in the usual
manner polarization is taken into account in classical electrodynamics? Another set
of questions arise about the radiational time arrow. The boundary 
conditions in time are realized on the microscopic level in QED and appear as
infinitesimal imaginary contributions in $S_Q$. How does the 
experimentally well established retarded Green function arise from the time 
reversal invariant dynamics? Finally, such a calculation addresses the
measurement theory of Quantum Mechanics. In fact, in analyzing the measurement 
process one always ends up with the study of the interactions between a small and a 
large system. Are there some peculiarities in the dynamics when the large 
parameter, the ratio of the number of degrees of freedom of the large and the small
system, diverges, like spontaneous symmetry breaking occuring in some models 
in the thermodynamical limit? We attempt to give below at least some indications about 
the answers. 

The CTP formalism has originally been introduced in Quantum Field Theory 
\cite{schw,jordan,mahanthappa,craig,keldysh,leplae,zhou,su,dewitt,arimitsu,umezawa,campos}
for the description of the time dependence of expectation values. 
Similar scheme was developed independently in the framework of Quantum Mechanics, too. 
The formal invariance of transition probabilities under time inversion lead to the 
construction of time symmetrical Quantum Mechanics \cite{ahare,ahark}.
The explicit appearance of the density matrix in the formalism explains the 
applications of this scheme in calculating the reduced density matrix in Quantum Mechanics 
\cite{feve} with special attention paid to dissipative processes
\cite{cale,zwer,haam,gasi,crv}. 
Furthermore, the description of decoherence \cite{decore,zureke,joze,decork}
in the framework of consistent histories \cite{histe,histk,histh}
can be achieved in a natural manner \cite{hpz,zph,phz} in the CTP formalism. 
Finally, one should mention promising applications of this scheme to 
cosmological problems \cite{cosm,cosme,cosmk,cosmh}, to kinetic Quantum Field 
Theory \cite{kine,kink,kinh,kinn} and to the renormalization group \cite{rge,rgk}.

In these applications of the CTP formalism the degrees of freedom are doubled,
due to the systematic implementation of time evolution in the Heisenberg picture 
and/or to the canonical description of dissipative processes. But this doubling 
remains rather formal from the point of view of classical physics.
The present work is based on the observation that one can actually construct
variational principle and canonical dynamics for the expectation values
of observables within the CTP
formalism. The choice of the microscopic degrees of freedom which become 
classical in the macroscopic region is unique up to a parameter which is
allowed due to the unitarity of the time evolution for closed systems.
The other, independent combination of the doubled degrees of freedom
takes care of the coupling of the system to its environment and becomes
suppressed in the macroscopic region. Such an explicit splitting of the
variables offers a new, more natural starting point for the exploration of the 
quantum-classical crossover, presented in this paper within the framework of QED.
The scope of this work is restricted to general, qualitative issues, such as
the current dynamics in the Dirac-see, the polarizability of the
vacuum, the radiation time arrow, the decoherence of non-relativistic charges
and the renormalization group scheme for the density matrix to describe
the quantum-classical crossover. The classical dynamics derived 
in these contexts yields new results already at such a qualitative
level. We plan to present the detailed, 
quantitative studies of these and related issues in subsequent publications.

The organisation of this paper is the following.
The action for the expectation values and some partial answer to these questions 
will be found within the CTP formalism, introduced in Section
\ref{exalqft} with special attention paid to the time evolution of the averages 
of observables and the role of the boundary conditions in time. The simplest system
where the role of the boundary conditions in time can be traced in forming the
time arrow is that of non-interacting particles. Section \ref{photons} contains the
derivation of the action for the expectation value of the free photon field. 
The electric current is a composite operator and its expectation value is controlled by a
substantially more complicated action. Section \ref{nondse} is devoted to these
complications and the presentation of the quadratic approximation to the action for the current. 
The case of interacting electrons and photons is considered in Section \ref{inteph}.
The use of classical expectation values, obtained with resolutions belonging to
the quantum domain, in particular decoherence, radiational time arrow and
vacuum polarization are commented on briefly in Section \ref{miclth}. The systematic 
study of the dependence of the expectation values on the space-time resolution
around the quantum-classical crossover can be achieved by means of the extension 
of the renormalization group strategy for the density matrix. Some qualitative remarks 
about the renormalized trajectory of this scheme are presented in Section \ref{qrg}. 
Finally, Section \ref{concl} is reserved for the summary of the results. The technical
parts of the calculation are collected in appendices. The two-point function
of a hermitian local operator of the CTP formalism is introduced in Appendix \ref{freegenf}
and some details of the calculation of the two-point vertex function in the non-interacting 
Dirac-see are collected in Appendix \ref{qefnds}.
In Appendix \ref{loopexp} the expressions needed for the connected two-point
functions for the current and the photon field are presented. The calculation of the 
effective action for the expectation values of these operators is outlined in Appendix \ref{qedgam}.
The evolution equation for the functional renormalization group in the CTP
formalism is derived in Appendix \ref{frgctp}.

\section{Expectation values in Quantum Field Theory}\label{exalqft}
Our goal is to establish relations among the expectation values of a set of
local observables like $\la\Psi(t)|O_a(\v{x})|\Psi(t)\ra_S$ given in the Schr\"odinger 
representation. The problem is interesting when the system is in an excited state.
The only technical restriction on the nature of the excitations is that the initial
state should be the result of an adiabatic time evolution generated from the ground state
by adding the source term $\sum_aj_a(t,\v{x})O_a(\v{x})$ to the Hamiltonian density
where the summation is over the observables considered. Closed systems will be 
considered below and the issue of relaxation and equilibrium in the presence
of a reservoir will not be addressed.
The state vector of the system $|\Psi(t)\ra_S=U(t,t_i)|\Psi_i\ra_S$
is given in terms of the initial condition $|\Psi_i\ra_S$ imposed at time $t_i$ and the 
time evolution operator $U(t,t_i)=\exp-i(t-t_i)H$ involving the Hamiltonian $H$.
Such expectation values are usually given in the Heisenberg representation,
$\la\Psi_i|A(x)|\Psi_i\ra_H$, where $x=(t,\v{x})$ and $A(x)=U(t_i,t)A(\v{x})U(t,t_i)$
and their perturbation series is obtained in terms of Green functions which are 
different from the usual ones occuring in the expressions of scattering 
amplitudes. Although the trajectory $|\Psi(t)\ra_S$ is fixed in the space of states 
by the initial condition $|\Psi_i\ra_S$ the superposition principle allows us to
project the system at time $t_f$ on a given final state $|\Psi_f\ra_S$
and to define the scalar product $\la\Psi_f|\Psi(t)\ra_S=\la\Psi_f|U(t_f,t_i)|\Psi(t)\ra_S$
interpreted as the transition amplitude between initial and final states. 
Its perturbative series contain matrix elements like
$\la\Psi_f|U(t_f,t)A(\v{x})U(t,t_i)|\Psi_i\ra_S$ which are not interpretable in 
terms of measurements according to standard rules of Quantum Mechanics. Such
matrix elements are reduced to an expectation value for the trivial case
$|\Psi_f\ra=|\Psi_i\ra=|0\ra$ only (by setting the ground state energy to zero).

Let us return to our problem, the calculation of the expectation values in excited states
and consider the generalization of the forward scattering amplitude for several observables,
\be\label{gfrtra}
\la\Psi_i|T[U(t_f,t_i)A_n(x_n)\cdots A_1(x_1)]|\Psi_i\ra_H
=\la\Psi_i|T[U(t_f,x^0_n)A_n(\v{x}_n)U(x^0_n,t_i)\cdots 
U(t_i,x^0_1)A_1(\v{x}_1)U(x^0_1,t_i)]|\Psi_i\ra_S
\ee
which differs from the Green function defined as the expectation value of the 
operator $T[A_n(x_n)\cdots A_1(x_1)]_H$, \cite{schw},
\be\label{gfhr}
\la\Psi_i|T[A_n(x_n)\cdots A_1(x_1)]|\Psi_i\ra_H
=\la\Psi_i|U(t_i,t_f)T[U(t_f,x^0_n)A_n(\v{x}_n)U(x^0_n,t_i)\cdots 
U(t_i,x^0_1)A_1(\v{x}_1)U(x^0_1,t_i)]|\Psi_i\ra_S.
\ee
The difference occurs in the way the scattering matrix $U(t_f,t_i)$ appears in these 
expressions. In Eq. \eq{gfrtra} the time evolution is constrained 
in such a manner that the system arrives at the state $|\Psi_i\ra$ at time $t_f$,
in contrast to Eq. \eq{gfhr} where the time evolution is open ended, without
any constraint on the evolution after the action of the observables. 
The functions given by Eqs. \eq{gfrtra} and
\eq{gfhr} are identical for $|\Psi_i\ra=|0\ra$ because the vacuum is stable
during the time evolution. But the dynamics of expectation values 
in an excited state, $|\Psi_i\ra\not=|0\ra$, requires the use of the 
expectation value \eq{gfhr} obtained in the closed time path (CTP) formalism
\cite{jordan,mahanthappa,craig,keldysh,leplae,zhou,su,dewitt,arimitsu,umezawa,campos}, 
rather than the contribution \eq{gfrtra} to the scattering amplitude.
The careful reader may object that it is sufficient to use the vacuum state
because all physical states can be obtained from the
vacuum by applying local excitations, ie. $|\Psi_i\ra_H=A_i(x_i)|0\ra_H$
where $A_i(x)$ is a local operator in space-time. Extend the time evolution
in this case in such a manner that the system starts at time $t=t_0<t_i$ and ends
at time $t=t_1>t_f$ with the vacuum state and write Eq. \eq{gfhr} as
\be
\la0|\at[A_i(x_i)U(t_0,t_1)]T[U(t_1,x^0_n)A_n(\v{x}_n)U(x^0_n,t_i)\cdots 
U(t_i,x^0_1)A_1(\v{x}_1)U(x^0_1,t_i)A_i(x_i)U(t_i,t_0)]|0\ra_S,
\ee
where $\at$ denotes anti-time ordering. The comparison of this expression with
\eq{gfrtra} shows that the basic difference between the CTP 
formalism and the usual scattering amplitudes is the presence of the anti-time 
ordered piece in the former case. The proper treatment of such matrix elements 
requires the introduction of independent time variables for the time ordered and 
the anti-time ordered operators.

To compare the expectation values obtained by constrained and open ended time 
evolutions we slightly generalize the CTP formalism and allow
that the system be described by the density matrices $\rho_i$ and $\rho_f$
in the initial and the final states, respectively. Furthermore, we introduce two sets 
of external sources, $j^\pm_a(x)$, coupled linearly to a number of local observables, 
$O_a(\v{x})$, in the hermitian Hamiltonian which is extended to
\be
H^\pm(t)=H\mp\sum_a\int d^3xj^\pm_a(t,\v{x})O_a(\v{x}).
\ee
Finally, we define the generating functional for the connected Green functions
\be\label{wdefe}
e^{iW[j^+,j^-]}=\Tr\at[e^{i\int_{t_i}^{t_f}dtH^-(t)}]\rho_fT[e^{-i\int_{t_i}^{t_f}dt'H^+(t')}]\rho_i
\ee
given in the Schr\"odinger representation. The sources play a double role
made explicit by the parameterization $j^\pm=j/2\pm\jb$. On the one 
hand, the variation of $j$ is used to generate the desired expectation values, 
and on the other, $\jb$ drives the system adiabatically from the vacuum at time 
$t=t_0$ to the desired initial state
$|\Psi_i\ra$ at $t=t_i$.  This allows us to set $\rho_i=|0\ra\la0|$ at $t_0=-\infty$ with 
the price of having non-vanishing physical external sources, $\jb$, for some initial 
times $t<t_i$. The quantity \eq{wdefe} is the transition probability between the 
states specified by the density matrices $\rho_i$ and $\rho_f$ for physical external 
sources. Expressions like \eq{wdefe} have already been used a number of times with $O_a$ 
chosen to be projection operator, eg. for the time-symmetric formulation of Quantum 
Mechanics \cite{ahare,ahark} and for the study of histories \cite{histe,histk,histh}.

It is natural to introduce the parameterization 
\be\label{fieldecomp}
\phi^\pm=\phi\pm\frac{\phi^\mr{adv}}{2}
\ee
for the field variables. The measured expectation values of the field at a given time $t$
involve the diagonal part of the functional $\la\phi^+|\rho_i(t)|\phi^-\ra$ with 
$\phi^\mr{adv}=0$. The canonical momenta, represented by the operators 
$\Pi^\pm=-i\delta/\delta\phi^\pm$, use a slightly extended domain, $\phi^\mr{adv}\approx0$,
of the density matrix.

There are two time axis in the CTP formalism and each degree of freedom 
exists in two copies, realized first in the case of thermal equilibrium,
$\rho_i=Z^{-1}\exp-H/T$, \cite{haag,arakie,arakik,lande,landk},
but can also be clearly seen in the path integral representation
\be\label{piwdefe}
e^{iW[j^+,j^-]}=\int D[\phi^+]D[\phi^-]e^{iS[\phi^+]-iS^*[\phi^-]+\sum_\sigma\int_{t_i}^{t_f}dt\sum_a
\int d^3xj_a^\sigma(t,\v{x})O_a(\phi^\sigma(t,\v{x}))]}]
\rho_f[\phi^-(t_f),\phi^+(t_f)]\rho_i[\phi^+(t_i),\phi^-(t_i)]
\ee
given as a functional integration over the trajectories $\phi^\sigma$, $\sigma=\pm$,
with the density matrix elements $\la\phi^+|\rho|\phi^-\ra=\rho[\phi^+,\phi^-]$ in the integrand. 
The duplication is made explicit by defining two kinds of averages, one for each time axis,
\be
\la\at[A],T[B]\ra^{j,\jb}_{\rho_f,\rho_i}=\Tr\at[Ae^{i\int_{t_i}^{t_f}dtH^-(t)}]
\rho_fT[Be^{-i\int_{t_i}^{t_f}dt'H^+(t')}]\rho_i.
\ee
Notice that $\la\openone,T[B]\ra^{0,\jb}_{\rho_f,\rho_i}$ is real according to the relation
\be
(\la\at[A],T[B]\ra^{j,\jb}_{\rho_f,\rho_i})^*=\la\at[B],T[A]\ra^{-j,\jb}_{\rho_f,\rho_i},
\ee
showing an inherent time inversion invariance ie. the time inversion leaves 
$\rho_i$, $\rho_f$ and $\jb$ invariant and flips the sign of the book-keeping variable 
$j$ only. The time-dependence of the density matrices
\bea
\rho_i(t)&=&T[e^{-i\int_{t_i}^tdt'H^+(t')}]\rho_i\at[e^{i\int_{t_i}^tdtH(t)}],\nn
\rho_f(t)&=&\at[e^{i\int_t^{t_f}dtH(t)}]\rho_fT[e^{-i\int_t^{t_f}dt'H^+(t')}]
\eea
allows us to write
\be\label{genprob}
e^{iW[j^+,j^-]}=\Tr\rho_f(t)\rho_i(t),~~~~~~(t_i\le t\le t_f).
\ee
The right hand side of this equation is the scalar product of the inital and final
hermitean density matrices taken at an arbitrary time indicating that the CTP formalism
is based on the transition probabilities rather than amplitudes and is time
reversal invariant. The duplication of the time
variables represents the independence of the quantum fluctuations
in the bras and the kets of the density matrices. We shall argue below 
that these fluctuations become correlated and the usual single time axis formalism 
is recovered in the macroscopic limit when decoherence \cite{decore,decork} 
suppresses the off-diagonal elements of the density matrix. 

It will be instructive to follow the dynamics in the presence of two different boundary 
conditions in time. The fixed boundary condition (FBC) for pure initial and final states, 
defined by $\rho_i=|\Psi_i\ra\la\Psi_i|$, $\rho_f=|\Psi_f\ra\la\Psi_f|$ 
decouples the dynamics of the two time axis. The additivity of transition probabilities
yields
\be
\la\openone,\openone\ra^{0,\jb}_{\rho^{(1)}_f+\rho^{(2)}_f,\rho_i}
=\la\openone,\openone\ra^{0,\jb}_{\rho^{(1)}_f,\rho_i}
+\la\openone,\openone\ra^{0,\jb}_{\rho^{(2)}_f,\rho_i}
\ee
for $\Tr\rho^{(1)}_f\rho^{(2)}_f=0$. The open boundary condition (OBC) defined by 
$\rho_i=|\Psi_i\ra\la\Psi_i|$ and $\rho_f=\sum_n|\Psi^{(n)}\ra\la\Psi^{(n)}|=\openone$,
$\{|\Psi^{(n)}\ra\}$ being a basis for the states corresponds to unconstrained
time evolution and couples the dynamics of the two time axis at $t=t_f$. Note that the time 
$t_f$ at which the final condition $\rho_f=\openone$ is imposed can be chosen
arbitrarily for unitary time evolution as long as it is later than the time for which
the latest observable inserted. We call a boundary condition reflecting if the time 
at which it is imposed as an initial or final condition influences the 
measurable expectation values. The open boundary condition is non-reflecting 
because $\rho_f=\openone$ commutes with all observables. 

The main virtue of the Heisenberg representation and the CTP formalism
in particular, is to render the initial condition problems of Quantum Mechanics 
simple. The hyperbolic Schr\"odinger equation allows us to solve the initial condition
problem in terms of state vectors but these are in general complicated objects. 
We can deal more efficiently with numbers, for instance matrix elements and therefore 
it is advantageous to convert operator equations into c-number equations. But a matrix 
element contains a bra and a ket, corresponding to the final and initial states, 
respectively as in Eq. \eq{gfrtra} and it is not clear how to express the solution of 
the initial condition problem of the Schr\"odinger equation with open ended time 
evolution in terms of such matrix elements. This is the problem which is solved in the 
framework of the CTP formalism by extending $\rho_i$ and $\rho_f$ to positive 
semidefinite hermitean operators beyond the domain of density matrices.

The direction of the time is determined by the phase of an eigenstate of the 
Hamiltonian as a function of the time, and the time runs in opposite directions along 
the two time axis. Due to the presence of both time directions there is no explicit 
time arrow in the expectation values which contain both retarded and advanced effects. 
But one can separate these effects, as far as the external sources are concerned, 
in the case of open boundary condition and we shall see below that 
$\phi$ and $\phi^\mr{adv}$ of Eq. \eq{fieldecomp} are the dynamical variables which are 
build up by retarded and advanced effects of the external sources, respectively.

An important property of the open boundary condition with unitary time evolution is that 
the generating functional is vanishing for physical external sources, $W[j,-j]=0$. 
As a result we have to vary non-physical sources, $j$, to generate the measured 
expectation values $\la O_a(x)\ra=\la\openone,O_a(x)\ra=\la O_a(x),\openone\ra$.
This suggests the slight generalization of the parameterization of the sources, $\jb\to\jb+\kappa j$, ie.
\be\label{jpar}
j^\pm=\frac{j}{2}(1\pm\kappa)\pm\jb,
\ee
giving
\be\label{kapav}
\la O_a(x)\ra=\fd{W[j,\jb]}{j_a(x)}_{|j=0}=\frac{1+\kappa}{2}\la\openone,O_a(x)\ra
+\frac{1-\kappa}{2}\la O_a(x),\openone\ra
\ee
for arbitrary choice of $\kappa$. The $\kappa$-independence of the expectation values
reflects a one dimensional symmetry of the physical sector of the CTP formalism.

The parameter $\kappa$ appearing here mixes the non-diagonal quantum fluctuations into
the expectation value of the observables. We need $\kappa\not=0$ ie. both kind of 
fluctuations, to derive variational equations of motion for the measured expectation values. 
This condition seems natural because both canonical variables are needed in the 
dynamics but the source $j$ is coupled only to the diagonal quantum fluctuations for 
$\kappa=0$ which are not sufficient to reconstruct the expectation values of the 
canonical momenta $\Pi$. The detailed argument goes as follows.
The action constructed for the field variable $\phi$ is obtained in terms of
connected Green functions. The $n$-point functions give rise to
$O(\phi^{n-1})$ terms in the equation of motion. The simplest linearized 
equation of motion arises from some connected two-point functions. These
two-point functions must include field variables at different times
to generate time evolution for the expectation values. Let us consider the
combination 
\be
G=\frac{1}{2\Delta t}\la\Psi_i|\phi(t,\v{x})[\phi(t,\v{y})-\phi(t-\Delta t,\v{y})]|\Psi_i\ra_H
\ee
of two-point functions with small $\Delta t$ as a typical term
which can also be written as 
\be
G=-\frac{i}{2}\la\Psi_i|\phi(t,\v{x})[\phi(t,\v{y}),H]|\Psi_i\ra_H+O(\Delta t)
\ee
according to Eq. \eq{gfhr}. The kinetic energy $\Pi^2(t,\v{x})/2$ in the Hamiltonian 
density yields
\be
G=\la\Psi_i|\phi(t,\v{x})\Pi(t,\v{y})|\Psi_i\ra_H+O(\Delta t),
\ee
and shows that the appearance of the canonical momentum operator is unavoidable in the 
Green functions involving fields at different times. We have the option of 
keeping in the formalism the expectation values of both canonical variables, 
$\phi$ and $\Pi$, the resulting variational equations being the quantum analogies
of the Hamilton equations. But once we have decided to retain the coordinate averages 
only we need $\kappa\not=0$ to couple both diagonal and off-diagonal 
fluctuations to the source $j$. The off-diagonal fluctuations drop out from the field
average \eq{kapav} but the $O(j^2)$ terms in the generating functional 
$W[j,\jb]$ and the $O(\phi^2)$ term of the action will retain them.

A simpler but more formal reasoning starts with the identity $W^*[j^+,j^-]=-W[-j^-,-j^+]$
obtained by comparing $W[j^+,j^-]$ with its complex conjugate in Eq. \eq{wdefe}. 
This identity reads as $W^*[j,\jb]=-W[-j,\jb]$
for $\kappa=0$, indicating that the real part of $W[j,\jb]$ which will be important 
for the equation of motion is an odd functional of $j$ and has, in particular an
$O(j^2)$ part. As a result the equation of motion for $\phi$ contains even powers
of the field variables. In order to have linear part in the equation of motion 
we have to allow $\kappa\not=0$.

As a simple demonstration that the measured expectation value
results from the retarded effects of the classical external sources 
we consider the linear response formulae. A physical external source, 
$j_+^a(x)=-j_-^a(x)=\jb^a(x)$ drives the time evolution of the expectation value of 
the operator $O_\ell(x)$ in the perturbation expansion of the external source 
according to
\bea
\la\openone,O_\ell(x)\ra_{\openone,\rho_i}^{0,\jb}&=&-i\fd{}{j^+_\ell(x)}
\sum_{n_+,n_-=0}^\infty\frac{(-1)^{n_-}}{n_+!n_-!}
\left(\sum_a\int dx'j^a(x')\fd{}{j^+_a(x')}\right)^{n_+}\nn
&&\times\left(\sum_b\int dx''j^b(x'')\fd{}{j^-_b(x'')}\right)^{n_-}
{e^{iW[j^+,j^-]}}_{|j^+=-j^-=\jb}.
\eea
The linear response formula for $O_\ell(t)$ is the $O(\jb)$ part of the right hand side,
\be
i\int dx'\sum_a\jb_a(x')\left(\la O_a(x'),O_\ell(x)\ra_{\openone,\rho_i}^{0,0}
-\la\openone,T[O_a(x')O_\ell(x)]\ra_{\openone,\rho_i}^{0,0}\right),
\ee
which can be written as
\bea\label{pabav}
&&i\int d^3x'\int_{t_i}^{t_f}dt'\sum_a\jb_a(t',\v{x}')\Theta(t'-t)
\left(\la O_a(x'),O_\ell(x)\ra_{\openone,\rho_i}^{0,0}
-\la\openone,T[O_a(x')O_\ell(x)]\ra_{\openone,\rho_i}^{0,0}\right)\nn
&&+i\int d^3x'\int_{t_i}^{t_f}dt'\sum_a\jb_a(t',\v{x}')\Theta(t-t')
\left(\la O_a(x'),O_\ell(x)\ra_{\openone,\rho_i}^{0,0}
-\la\openone,T[O_a(x')O_\ell(x)]\ra_{\openone,\rho_i}^{0,0}\right).
\eea
If the final time boundary condition is non-reflecting and can be imposed at any time 
superior to $t$ without modifying the expectation values then the relation 
\be\label{circid}
\Theta(t'-t)\la O_a(x'),O_\ell(x)\ra_{\openone,\rho_i}^{0,0}
=\Theta(t'-t)\la\openone,T[O_a(x')O_\ell(x)]\ra_{\openone,\rho_i}^{0,0}
\ee
holds. In fact, one can reduce $t_f$ down to $t'$ followed by the commutation
of the perturbation $O_a(x')$ with $\rho_f$ and the placing of it at the end of the 
anti-time ordered part of the expression. This identity shows the
cancellation of the advanced part of the causal propagator and explains that the usual
retarded response formulae are the result of the interference between the dynamics 
evolving along the two different time paths. Whenever a final boundary condition 
is used which can not be displaced in time
because it does not commute with all operators in question then advanced effects are
left behind. Notice that this remains valid in any order. In fact,
the possibility of reducing $t_f$ until it reaches the time of observation,
$t$, removes any influence of the sources on the expectation value after
the time of the measurement. The condition of the cancellation was that the 
non-vanishing sources are classical, $j^+=-j^-$. The non-classical part of the source, 
$j$, representing a coupling among the degrees of freedom of a closed quantum system may 
induce advanced effects. But this component is suppressed in the macroscopic limit by decoherence.

\section{Photons}\label{photons}
The simplest context in which the dynamics of the field expectation values can be studied is
the case of free photons. The expectation values are obtained in two steps. First
the generating functional for the connected Green functions of the conserved current 
is calculated. This produces the expectation values in terms of the
external sources which drive the system to the desired initial condition.
In the second step a number of different effective actions and their variational 
equations of motion are constructed by performing a Legendre transformation on 
the external sources.

\subsection{Photon propagator}
The generating functional for the connected Green functions is constructed by
coupling an external current to the photon field and using the pair of extended
Hamiltonians $H\to H^\pm(t)=H\mp\int d^3xj^{\pm\mu}(t,\v{x})A_\mu(\v{x})$ in
Eq. \eq{wdefe}. It is advantageous to carry out the calculations in the path 
integral formalism where the action for the photon field is written as
\bea
S_M[A]&=&\int_x\left[-\frac14(\partial_\mu A_{\nu,x}-\partial_\nu A_{\mu,x})^2
-\frac{\xi}{2}(\partial^\mu A_{\mu,x})^2\right]\nn
&=&\hf A\cdot D^{-1}_0\cdot A
\eea
Here $D_0$ is the free photon propagator $D^{-1}_0=D^{-1}_T+D^{-1}_L$
and the projection operators onto the transverse and longitudinal components of the 
photon field $T^{\mu\nu}=g^{\mu\nu}-L^{\mu\nu}$ and $L^{\mu\nu}=\partial^\mu\partial^\nu/\Box$
are used to construct $D^{-1}_T=(\Box-i\epsilon)T$ and $D^{-1}_L=\xi(\Box-i\epsilon)L$. 

A few words about the notations: The field configurations $\phi(x)$ are usually considered 
as vectors, $\phi_x$, and space-time integrals with
occasional summation over repeated indices as scalar products, eg.
$\int dx\phi(x)\chi(x)=\int_x\phi_x\chi_x=\phi\cdot\chi$ or
$\int dxA_\mu(x)j^\mu(x)=A\cdot j$. The space-time coordinates are written as
$x=(t,\v{x})$, the space-time indices are given by Greek letters, the
Latin letters denote combined indices like $a=(\mu,x)$. Repeated indices are summed/integrated
over.

The generating functional $W^\mr{phot}[j^+,j^-]$ is given in the framework of 
the path integral representation as
\be\label{wpi}
e^{iW^\mr{phot}[\hj]}=\int D[\hA]e^{\frac{i}{2}\hA\cdot\hD^{-1}_0\cdot\hA+i\hj\cdot\hA},
\ee
where the two component integral variables $\hA=(A^+,A^-)$ and external sources
$\hj=(j^+,j^-)$ are introduced to simplify the expressions. 
The current $j^\pm$ will be parameterized as in Eq. \eq{jpar}.
Some complications arise due to the presence of the density matrices $\rho_i$ and
$\rho_f$ in the functional integral. Anticipating the perturbation expansion
we use Gaussian density matrices which generate quadratic expressions in the
action of the path integral, ie. $\ln\rho_i[A_{+,t_i};A_{-,t_i}]$
and $\ln\rho_f[A_{-,t_f};A_{+,t_f}]$ should be at most quadratic in the fields.
The inverse block propagator is
\be\label{proppph}
\hD^{-1}_0=\begin{pmatrix}D_0^{-1}&0\cr0&-D_0^{-1*}\end{pmatrix}+\hD^{-1}_\mr{BC},
\ee
where the second term on the right hand sides stands for the contributions of the
density matrices and is non-vanishing for $t=t_i$ or $t_f$ only.
Notice the presence of an infinitesimal imaginary part in the inverse propagator
\eq{proppph}. The operator expression \eq{wdefe} contains a hermitian Hamiltonian because
the (anti)time ordering implies the time boundary conditions instruction. In contrast,
the time boundary conditions involve the infinitesimal imaginary parts in the
inverse propagators in the path integral representation.

Most of the work reported below is based on OBC with the vacuum as initial state.
Although one can determine $\hD_0^{-1}$ directly in the path integral 
formalism for this boundary condition it is simpler to construct 
the photon propagator, $\hD_0$, within the operator formalism. The result is
\be\label{photpr}
\hD_0=\begin{pmatrix}D_0^\mr{near}+i\Im D_0&-\hf D_0^\mr{far}+i\Im D_0\cr\hf D_0^\mr{far}+i\Im D_0&-D_0^\mr{near}+i\Im D_0\end{pmatrix}
\ee
in the base $(j^+,j^-)$ where $D^\mr{near}$ and $D^\mr{far}$ denote the near and far field Green functions
\cite{dirac}, respectively
and $i\Im D_0$ stands for the imaginary part of the causal (Feynman) Green function.
The retarded and advanced Green functions, $D_0^\mr{ret}=D_0^\mr{near}+\hf D_0^\mr{far}$ and 
$D_0^\mr{adv}=D_0^\mr{near}-\hf D_0^\mr{far}$ will frequently be used, as well, cf. Appendix \ref{freegenf}. 

The Gaussian integral can easily be carried out in Eq. \eq{wpi} yielding
\be\label{photw}
W^\mr{phot}[\hj]=-\hf\hj\cdot\hD_0\cdot\hj
\ee
and a straightforward calculation gives
\be\label{phw}
W_\mr{OBC}^\mr{phot}[j,\jb]=-\hf^{(\jb,j)}\cdot
\begin{pmatrix}0&D_0^\mr{adv}\cr D_0^\mr{ret}&\kappa D_0^\mr{near}+i\Im D_0\end{pmatrix}\cdot
\begin{pmatrix}\jb\cr j\end{pmatrix}
\ee
for open boundary condition. In the case of fixed boundary condition
the two time axis decouple yielding
\bea\label{phwf}
W_\mr{FBC}^\mr{phot}[\hj]&=&-\hf j^+\cdot(D_0^\mr{near}+i\Im D_0)\cdot j^++\hf j^-\cdot(D_0^\mr{near}-i\Im D_0)\cdot j^-\nn
&=&-\hf^{(\jb,j)}\cdot\begin{pmatrix}2i\Im D_0&D_0^\mr{near}+\kappa i\Im D_0\cr
D_0^\mr{near}+\kappa i\Im D_0&\kappa D_0^\mr{near}+\frac{1+\kappa^2}{2}i\Im D_0\end{pmatrix}\cdot
\begin{pmatrix}\jb\cr j\end{pmatrix}.
\eea

\subsection{Effective action}
The role of the effective action is to provide the functional for the
variational equations satisfied by the space-time dependent expectation values
and it is introduced by performing a functional Legendre transformation on 
the generating functional of connected Green functions. Special attention 
must be payed to the fact that the Legendre pair of the sources, the expectation
values, are complex in general. Accordingly we perform the Legendre transformation 
separately for the real and imaginary part of $W[\hj]=\Re W[\hj]+i\Im W[\hj]$. 
The measured expectation value of the photon field is always real and it will be
determined by $\Re W[\hj]$. Therefore, we start with the effective action
\be\label{phreg}
\Gamma^\mr{phot}[A,A^\mr{adv}]=\Re W^\mr{phot}[j,\jb]-\jb\cdot A^\mr{adv}-j\cdot A,
\ee
with independent variables consisting of the measured expectation value of the
photon field
\be
A=\fd{\Re W^\mr{phot}[j,\jb]}{j},
\ee
and an auxiliary field variable 
\be
A^\mr{adv}=\fd{\Re W^\mr{phot}[j,\jb]}{\jb}.
\ee
The inverse Legendre transform based on the relations \eq{phreg}
\bea\label{ivtrre}
j&=&-\fd{\Gamma^\mr{phot}[A,A^\mr{adv}]}{A},\nn
\jb&=&-\fd{\Gamma^\mr{phot}[A,A^\mr{adv}]}{A^\mr{adv}}
\eea
serves as equations of motion for the expectation values. Since the averages
$A$ and $A^\mr{adv}$ receive contributions from the diagonal and non-diagonal fluctuations, respectively,
these equations of motion control the time dependence for both the coordinates and the 
momenta.

The generating functional \eq{phw} yields
\be\label{regamgam}
\Gamma_\mr{OBC}^\mr{phot}[A,A^\mr{adv}]=-A^\mr{adv}\cdot D_0^{\mr{ret}-1}\cdot A
+\frac{\kappa}{2}A^\mr{adv}\cdot D_0^{\mr{ret}-1}\cdot D_0^\mr{near}\cdot D_0^{\mr{adv}-1}\cdot A^\mr{adv},
\ee
for open boundary condition. The corresponding equations of motion for $A$ and $A^\mr{adv}$ are
\be\label{eqmaakap}
A^\mr{adv}=D_0^\mr{adv}j,
\ee
and
\be\label{eqmakap}
A=D_0^\mr{ret}\jb+\kappa D_0^\mr{near}j,
\ee
respectively. They show that non-diagonal fluctuations contribute to the 
advanced field $A^\mr{adv}$ generated by $j$ and the physical
expectation value, $A$, is indeed the retarded field created by the physical external
current $\jb$. The two fields, $A$ and $A^\mr{adv}$ remain decoupled in the absence of interactions.
For fixed boundary conditions, Eq. \eq{phwf}, we have
\be\label{regamgamf}
\Gamma_\mr{FBC}^\mr{phot}[A,A^\mr{adv}]=-A^\mr{adv}\cdot D_0^{\mr{near}-1}\cdot A+\frac{\kappa}{2}A^\mr{adv}\cdot D_0^{\mr{near}-1}\cdot A^\mr{adv}.
\ee
This expression can be obtained from Eq. \eq{regamgam} by the replacements 
$D_0^\mr{ret}\to D_0^\mr{near}$ and $D_0^\mr{adv}\to D_0^\mr{near}$ which represent the
loss of the interference between the two time axis.

The effective actions introduced so far provide the equations of motion for the 
expectation values for both field variables of the CTP formalism. But the 
field $A^\mr{adv}$ is not physical and its presence is not necessary to extract the time
dependence of the photon field $A$. We simplify the Legendre transformation 
by keeping $\jb$ as a fixed parameter. The resulting effective action for the 
photon field alone is
\be\label{effgphd}
\Gamma^\mr{phot}[A]=\Re W^\mr{phot}[j,\jb]-j\cdot A,
\ee
where the dependence of $\Gamma^\mr{phot}[A]$ on $\jb$ is not shown explicitely.
The effective action for OBC turns out to be
\be\label{idefaa}
\Gamma_\mr{OBC}^\mr{phot}[A]=-\frac{1}{2\kappa}(A-\jb D_0^\mr{adv})\cdot D_0^{\mr{near}-1}\cdot(A-D_0^\mr{ret}\jb),
\ee
generating an equation of motion $\delta\Gamma^\mr{phot}[A]/\delta A=-j$ 
identical to Eq. \eq{eqmakap}. 

So far the value of the parameter $\kappa$ entering by expression \eq{jpar} is 
arbitrary. The choice $\kappa=0$ seems to be natural and simple 
and the corresponding currents $\jb$ and $j$ generate retarded or 
advanced Li\`enard-Wiechert potentials, respectively, according to the equations of 
motion. But the effective action \eq{regamgam} contains no $O(A^{a2})$ 
piece when $\kappa=0$ in agreement with the general remark made in Section \ref{exalqft}
stating that the non-physical field can not be eliminated and no classical action and 
variation principle can be found for the measured expectation value of the photon field. 
This is reflected in the appearance of the coefficient $1/\kappa$ in the effective 
action \eq{effgphd} because for $\kappa=0$ the generating functional \eq{phw} has no 
$O(j^2)$ term in the real part and the Legendre transformation is not defined 
for linear functions.

For OBC the Legendre transform of the imaginary part is defined as
\be\label{phimg}
\Gamma_\mr{OBC}^\mr{phot~im}[A^\mr{im}]=\Im W_\mr{OBC}^\mr{phot}[j]-j\cdot A^\mr{im},
\ee
where
\be
A^\mr{im}=\fd{\Im W^\mr{phot}_\mr{OBC}[j]}{j}.
\ee
The corresponding equation of motion is
\be\label{ivtrrei}
j=-\fd{\Gamma_\mr{OBC}^\mr{phot~im}[A^\mr{im}]}{A^\mr{im}}.
\ee
The generating functional \eq{phw} yields the explicit form
\be\label{imaobc}
\Gamma_\mr{OBC}^\mr{phot~im}[A^\mr{im}]=-\hf A^\mr{im}\cdot\Im D_0^{-1}\cdot A^\mr{im}.
\ee

For FBC the functional $\Im W[\hj]$ depends on both currents and the Legendre 
transformation results in
\be\label{imafbc}
\Gamma_\mr{FBC}^\mr{phot~im}[A^\mr{im},A^\mr{adv~im}]=\Im W^\mr{phot}_\mr{FBC}[j,\jb]-j\cdot A^\mr{im}-\jb\cdot A^\mr{adv~im}
\ee
where
\bea
A^\mr{im}&=&\fd{\Re W^\mr{phot}_\mr{FBC}[j,\jb]}{j},\nn
A^\mr{adv~im}&=&\fd{\Re W^\mr{phot}_\mr{FBC}[j,\jb]}{\jb}.
\eea
The actual form,
\be
\Gamma_\mr{FBC}^\mr{phot~im}[A^\mr{im},A^\mr{adv~im}]=-\frac{1+\kappa^2}{4}A^\mr{adv~im}\Im D_0^{-1}A^\mr{adv~im}-A^\mr{im}\Im D_0^{-1}A^\mr{im}+\kappa A^\mr{im}\Im D_0^{-1}A^\mr{adv~im},
\ee
follows from the functional \eq{phwf}.

These effective actions will be used in Section \ref{decohnrc}.

\section{Current dynamics in the non-interacting Dirac-see}\label{nondse}
For a non-interacting system the determination of the expectation value of a local 
observable is trivial as long as the observable is a one-body operator,
as in the case of the photon field discussed in the previous Section.
The triviality comes from the fact that the generating functional for the connected 
Green function of the elementary fields is quadratic and can exactly be calculated. 
Higher order Green functions,
$\la0|T[\phi_{x_1}\cdots\phi_{x_{2n}}]|0\ra$ with $n\ge2$, factorize to a sum of
$n$ disconnected products of two-point functions, according to Wick-theorem.
The physical origin of this factorization is obvious, it comes from the absence
of interactions between the particles. Each particle created by one of the 
operators must be destroyed by another operator in order to find a non-vanishing 
contribution to the vacuum expectation value. In other words, the
largest cluster with non-factorizable structure contains two operators.
But we meet serious difficulties as soon as the expectation value of such an
operator is sought which controls more than one particle. The non-factorization
arises from the possibility of creating particles by one operator which are
annihilated by different other operators. One has to perform a "bosonisation"
in a fermionic system because only bosonic operators can have 
non-vanishing expectation values. The simplest and most important bosonic operator 
is the electric current $j$ considered in this Section.

Let us first introduce the generating functional, $W^\mr{el}[\ha]$, for the connected Green 
functions of the electric current by means of the functional 
\be\label{gfnids}
e^{iW^\mr{el}[\ha,\hta,\htab]}=\int D[\hsi]D[\hsib]e^{i\hsib\cdot[\hG^{-1}_0+\ha\bre]\cdot\hsi
+i\htab\cdot\hsi+i\hsib\cdot\hta+iS^e_{CT}[\ha]}
\ee
where the two-component fields $\hsi=(\psi_+,\psi_-)$ $\ha=(a^+,a^-)$
were introduced together with the inverse electron propagator
\be
\hG^{-1}_0=\begin{pmatrix}G^{-1}_0\cr0&-\gamma^0G^{-1\dagger}_0\gamma^0\end{pmatrix}
+\hG_\mr{BC}^{-1},
\ee
where $G_0^{-1}=i\partial\br-m$. We shall need later the composite operator counterterm
\be
S^e_{CT}[\ha]=\frac{\Delta Z_3+\beta}{2}\ha\cdot
\begin{pmatrix}D_T^{-1}&0\cr0&-D_T^{-1*}\end{pmatrix}\cdot\ha
\ee
with $\Delta Z_3$ being UV divergent and the finite part $\beta$ being fixed by
a renormalization condition.

In the absence of charges in the initial and final states we identify the generating functional
as $W^\mr{el}_0[\ha]=W^\mr{el}[\ha,0,0]$. When the system contains $n^-$ electrons and
$n^+$ positrons we take $\rho_i=|\Psi_i\ra\la\Psi_i|$ or $\rho_f=|\Psi_f\ra\la\Psi_f|$,
\be
|\Psi\ra=\prod_{j=1}^{n^-}\left(\int_\v{y}\psib_{t,\v{y}}\chi^-_{j,\v{y}}\right)
\prod_{k=1}^{n^+}\left(\int_\v{y}\bar\chi^+_{k,\v{y}}\psi_{t,\v{y}}\right)|0\ra,
\ee
with $t=t_i$ or $t_f$, respectively. The wave functions $\chi^-$ and $\bar\chi^+$
describe the one-particle $e^\pm$ states. The generating functional for the current is 
then written as
\be
e^{iW^\mr{el}[\ha]}=\prod_{\sigma=\pm1}
\prod_{j=1}^{n^-}\left(\int_\v{x}\bar\chi^-_{j,\v{x}}\fd{}{\etab^\sigma_{t_f,\v{x}}}
\int_\v{y}\chi^-_{j,\v{y}}\fd{}{\eta^\sigma_{t_i,\v{y}}}\right)
\prod_{k=1}^{n^+}\left(\int_\v{x}\chi^+_{k,\v{x}}\fd{}{\eta^\sigma_{t_f,\v{x}}}
\int_\v{y}\bar\chi^+_{k,\v{y}}\fd{}{\etab^\sigma_{t_i,\v{y}}}\right)
{e^{iW^\mr{el}[\ha,\hta,\htab]}}_{|\hta=\htab=0}.
\ee

We want to generate the expectation value of the current
\be
j^\mu_x=\hf[\psib_x\gamma^\mu\psi_x-(\gamma^\mu\psi_x)^\mr{tr}\psib^\mr{tr}_x]
\ee
which changes sign under charge conjugation. To this end we use the $\gamma$-matrices
\be
(\gamma^\mu_x)_{y,z}=\hf(\delta_{y,x+\eta e^\mu}\delta_{z,x-\eta e^\mu}
+\delta_{y,x-\eta e^\mu}\delta_{z,x+\eta e^\mu})\gamma^\mu,
\ee
where $\eta=0^+$ in the Dirac Lagrangian and the minimal coupling. 
In this notation the current is $j_a=\psib\cdot\gamma_a\cdot\psi$.

The mathematical source of the complications is that
\be\label{wnisum}
W^\mr{el}[\ha]=-i\Tr\ln\hG^{-1}[\ha]+S_{CT}[\ha]
\ee
where $\hG^{-1}[\ha]=\hG^{-1}_0+\ha\br$ is an involved functional even for the vacuum, ie. 
in the absence of initial or final charges. It is easy to understand the origin of 
this complication in the special 
case of $\rho_i=\rho_f=|0\ra\la0|$ and time-independent source $\ha$. The generating functional 
$W^\mr{el}[\ha]$ is now the difference of the energy of the Dirac-see in the presence of the 
vector potential $a^+$ and $a^-$. A local modification of the vector potentials 
creates a non-local polarization of the Dirac-see with all filled one-electron states 
contributing. Therefore, the infinitely many negative energy states filled up in the 
Dirac-see renders the functional highly non-trivial. It is not necessary that
the states be filled. For scalar particles the structure of the generating functional
remains the same except of the change of the overall sign. The external potential
coupled to the particles creates a polarization of the ground state which involves an
arbitrary number of (non-interacting) particles. The expectation value of bilinear 
operators measures the polarization created by the given boundary condition in time and 
its space-time dependence satisfies highly non-trivial equations reflecting
the multi-particle dynamics of polarization in the ground state. 

Our strategy followed in the case of the non-interacting Dirac-see will be
similar to that of Section \ref{photons}, ie. we calculate first the generating
functional for the connected Green functions of the current and then proceed with the
construction of the effective actions. For weak fields the functionals can easily be 
calculated in the framework of the perturbation expansion, which is the main task
of this Section. The issue of strong fields is far more difficult and will only
be commented on briefly.

\subsection{Connected Green functions}
We start with the simpler case when there are no additional charges immersed in the
Dirac-see. For weak external sources the functional \eq{wnisum} can be written
as a functional Taylor series,
\be\label{waetaylor}
W^\mr{el}[\ha]=\sum_{n=1}^\infty\frac{1}{n!}W^\mr{el}_{0~a_1,\ldots,a_m}\ha_{a_1}\cdots\ha_{a_m}
\ee
where the super-index $a=(\pm,x,\mu)$ identifies a time axis, a space-time location 
and a vector index, and the coefficients give the connected Green functions.
The expansion of the logarithmic function in Eq. \eq{wnisum} results in
\be
W^\mr{el}[\ha]=-i\Tr\ln\hG^{-1}_0+i\sum_{n=1}^\infty\frac{(-1)^n}{n}\Tr(\hG_0\cdot\ha\br)^n+S_{CT}[\ha]
\ee
and
\be\label{saaa}
W^\mr{el}_{a_1,\ldots,a_n}=\frac{i(-1)^n}{n}\sum_{P\in S_n}\Tr[\hG_0\cdot\gamma^s_{\tilde a_{P(1)}}\cdot\hG_0
\cdot\hat\gamma^s_{\tilde a_{P(2)}}\cdots\hG_0\cdot\hat\gamma^s_{\tilde a_{P(n)}}
\cdot\hG_0\cdot\hat\gamma^s_{\tilde a}]-\delta_{n,2}(\Delta Z_3+\beta)\hD_T
\ee
where
\be
\hat\gamma^a_x=\begin{pmatrix}\gamma^a_x&0\cr0&\gamma^a_x\end{pmatrix}.
\ee
The symmetrization with respect to the exchange of the external legs is achieved
by the summation over the permutations of the vertices in \eq{saaa}. The odd orders are 
vanishing according to Furry's theorem. For the sake of simplicity we truncate the 
generating functional at the quadratic order and write
\be\label{qnidsgf}
W^\mr{el}[\ha]=-\hf\ha\cdot\htG_R\cdot\ha
\ee
where the renormalized current two-point function $\htG_R=\htG_0-\beta\hD_T$ is given by
\bea\label{tgbev}
\tG^{++}_{0a,b}&=&-i\Tr[\hG^{++}_0\cdot\gamma_a\cdot\hG^{++}_0\cdot\gamma_b]+\Delta Z_3T(\Box-i\epsilon)\nn
\tG^{-+}_{0a,b}&=&-i\Tr[\hG^{+-}_0\cdot\gamma_a\cdot\hG^{-+}_0\cdot\gamma_b]\nn
\tG^{--}_{0a,b}&=&-i\Tr[\hG^{--}_0\cdot\gamma_a\cdot\hG^{--}_0\cdot\gamma_b]-\Delta Z_3T(\Box+i\epsilon).
\eea
The notation $\tG=\tG^{++}$ will be used in the rest of the paper.
Well known results (eg. Ref. \cite{itzu}) include the renormalized two point function
\bea\label{gradetg}
\tG^{\mu\nu}_{0q}&=&\int_xe^{-iqx}\tG^{\mu\nu}_{x,0}\nn
&=&T^{\mu\nu}\left[\frac{1}{15\pi}\frac{q^4}{m^2}+O\left(\frac{q^6}{m^4}\right)\right]
\eea
obtained in the framework of the gradient expansion where $\Im\tG=0$ since the creation of 
a mass-shell $e^-e^+$ pair is forbidden. We shall use the same parameterization of the propagator
\be
\htG_0=\begin{pmatrix}\tG^\mr{near}_0+i\Im\tG_0&-\hf\tG^\mr{far}_0+i\Im\tG_0\cr\hf\tG^\mr{far}_0+i\Im\tG_0&-\tG^\mr{near}_0+i\Im\tG_0\end{pmatrix}.
\ee
as for photons and the retarded and advanced current Green functions
will be defined by $\tG_0^\mr{ret}=\tG_0^\mr{near}+\hf\tG_0^\mr{far}$ and $\tG_0^\mr{adv}=\tG_0^\mr{near}-\hf\tG_0^\mr{far}$,
respectively.

An interesting feature of Eq. \eq{saaa} is the need of renormalization. 
The Green functions $W^\mr{el}_{0~a_1,\ldots,a_n}$ are finite for $n\ge3$ but the 
two-point function diverges. In fact, this two-point function is identical 
to the one-loop photon self energy except of the missing factor $e^2$. 
The electrons of the Dirac-see tend to
approach each others too frequently and make the two point function
divergent when the two legs approach each other in space-time.
This is a well known problem in QED and the cure is the introduction of the counterterm
$S_{CT}[\ha]$. The UV finite part, $\beta$, of the counterterm is fixed by a 
renormalization condition to be imposed. The lesson of this divergence,
a relativistic effect, is that the dynamics of the current $j$ can
not be defined without an additional scale, the cutoff, even in the absence 
of interactions.

The need of a renormalization condition for non-interacting particles demonstrates 
a characteristic difference between first and second quantized systems. In quantum 
mechanics observables are defined by the operators. In quantum field theory however the
observables represented by composite operators may need counterterms and their proper 
definition must include the corresponding renormalization condition. Symmetry 
principles can not fix the counterterm to the current as long as it is transverse 
and we find a one-parameter family of current with the two-point function
\be\label{betan}
i\la0T[j_x^\mu j_{x'}^\nu]0|\ra+\beta\Box\delta_{x,x'}T^{\mu\nu},
\ee
as far as the many-body aspects are concerned. The finite part of the 
counterterm influences the product of two current operators at zero separation and
this contact term plays an important role even at finite energies. We shall see 
later that the current coupled to the photon field is defined by $\beta=0$.

The generating functional is more complicated in the presence of additional charges
in the Dirac-see. We then have the additional term $\htab\cdot\hG[\ha]_0\cdot\hta$ 
in $W[\ha,\htab,\hta]$ which gives for instance 
\be\label{egyel}
W^\mr{el}[\ha]=W_\mr{FBC}^e[\ha]-i\sum_{\sigma=\pm}\ln(\hG[\ha]^{\sigma\sigma}_{(j_f,x_f),(j_i,x_i)})
\ee
for charged fixed boundary condition with a single electron.
The initial (final) state is characterized by the space-time point 
$x_i$ ($x_f$) with bispinor index $j_i$ ($j_f$) and $a=(x,j)$ denotes the combined 
index. As an other example, the open boundary condition for two electrons with
$a_n$, $n=1,2$ in the initial state yields
\be\label{ketel}
W^\mr{el}[\ha]=W_\mr{OBC}^e[\ha]-i\ln(\hG[\ha]^{+-}_{a_1a_1}\hG[\ha]^{+-}_{a_2a_2}
-\hG[\ha]^{+-}_{a_1a_2}\hG[\ha]^{+-}_{a_2a_1}).
\ee
Now the generating functional contains odd orders in $\ha$. The different 
structure of the see and valence contributions, the first and the second terms
on the right hand side of Eqs. \eq{egyel}-\eq{ketel}, reflects the 
fact that the Dirac-see is made up of negative energy one-particle states 
while the additional charges, introduced by the creation operators, 
correspond to positive energy. As a result, the valence charges 
move freely while the motion of the particles making up the Dirac-see is
restricted by the Pauli-blocking.

\subsection{Effective actions}
The equation of motion for the current expectation values is derived from
the effective action
\be
\Gamma^\mr{el}[J,J^\mr{adv}]=\Re W^\mr{el}[a,\ab]-\ab\cdot J^\mr{adv}-a\cdot J
\ee
involving the sources
\be\label{apar}
a^\pm=a\frac{1\pm\kappa}{2}\pm\ab
\ee
and independent variables
\bea
J&=&\fd{\Re W^\mr{el}[a,\ab]}{a},\nn
J^\mr{adv}&=&\fd{\Re W^\mr{el}[a,\ab]}{\ab}
\eea
which are conserved currents.

The effective action for the physical field $A$ only is defined by
\be
\Gamma^\mr{el}[J]=\Re W^\mr{el}[a,\ab]-a\cdot J.
\ee
The actual expressions for the Dirac-see can be obtained by the replacements
$D_0\to\tG$, $A\to J$, and $A^\mr{adv}\to J^\mr{adv}$ in Eqs. \eq{regamgam}, \eq{regamgamf} and \eq{idefaa}.

When charges are added to the Dirac-see the valence propagator contributes to the 
$\ha$-dependence in $W^\mr{el}[\ha]$. This requires $\hj\not=0$ and a redefinition of 
$\htG$ in Eq. \eq{qnidsgf} to be used in the Legendre transformation.

\subsection{Polarized charges}\label{wpres}
Let us consider now the dynamics of localized charges polarized from the Dirac-see
and described by the effective action truncated at the quadratic order,
\be\label{qfreact}
\Gamma^\mr{el}[J]=\frac{1}{2\kappa}J\cdot\Gamma^{(2)\mr{el}}\cdot J.
\ee
The current is supposed to be slowly varying over distances $\ell<1/m$ rendering the
leading order gradient expansion \eq{gradetg}
\be\label{qfreactk}
\Gamma^{(2)\mr{el}}=-\frac{T}{\frac{1}{15\pi m^2}(\Box-i\epsilon)^2-\beta(\Box-i\epsilon)}
\ee
applicable. Feynman's $\epsilon$-prescription is displayed in the expression 
explicitely for the calculation of the inverse of the kernel. This expression 
obtained for OBC holds for FBC, as well, as long as the total 
charge is vanishing and $J^\mu$ consists of closed flux-tubes. The calulation
indicated briefly in Appendix \ref{qefnds} for the choice $\beta=0$, justified for
the electric current in Section \ref{ingfct}, gives
\be\label{vfctrt}
\Gamma^{(2)\mr{el}\mu\nu}_{\ \ \ x,x'}=-\frac{15m^2}{8}\Theta((x-x')^2)T^{\mu\nu},
\ee
ie. the vertex function is step function, assuming the value $0$ or 
$-15m^2/8$ for spacelike or timelike separations, respectively.

Let us now assume the form
\be\label{curr}
J^\mu_x=g^{\mu0}\sum_jN_j\rho(|\v{x}-\v{x}_j|)
\ee
for the current where $N_j$ denotes the number of electrons of a localized state and $\int_\v{x}\rho(r)=1$. 
Notice that the charge is polarized out from the negative energy one-particle
states for weak external source and there is no reason for $N_j$ to be integer. 
The average distance between two world-tubes is given by
\be\label{distwt}
r_{jk}=\int_{\v{y},\v{z}}\rho(|\v{y}-\v{x}_j|)|\v{y}-\v{z}|\rho(|\v{z}-\v{x}_k|).
\ee

To identify the total mechanical energy we need $E_\mr{tot}=-\Gamma^\mr{el}_{0~\mr{OBC}}[J]/(t_f-t_i)$ 
in the limit $t_f-t_i\to\infty$,
\be
E_\mr{tot}=\frac{15m^2}{8}\biggl[-(t_f-t_i)N_\mr{tot}^2+\sum_{j,k}N_jN_kr_{jk}\biggr]
\ee
$N_\mr{tot}=\sum_jN_j$ being the total charge. The effective action for $\beta=0$ is 
IR finite for neutral systems only. This is a trivial result, reflecting the 
impossibility of polarizing out a net charge by a neutral external source. 

It is interesting to note that the action of the current defined by $\beta<0$ is 
that of a classical electrodynamical action with charge $g=1/\sqrt{-\beta}$
and electric susceptibility $O(g^2k^2/m^2)$, cf. Section \ref{vacpol} yielding
\be
E_\mr{tot}=-\frac{g^2}{8\pi}\sum_{j,k}N_jN_k\int_{\v{x},\v{y}}\rho(|\v{x}-\v{x}_j|)
\frac{1-e^{-\sqrt{15\pi}|\v{x}-\v{y}|mg}}{|\v{x}-\v{y}|}\rho(|\v{y}-\v{x}_k|).
\ee
The screening effect of vacuum polarization detected by the current defined with 
$\beta<0$ removes the IR divergence and allows a net charge.

Though it is perplexing to find interactions between localized charges in the 
non-interacting Dirac-see there is a simple explanation for it. 
In fact, let us consider the normalized two-particle state 
\be
\psi(\v{x}_1,\v{x}_2)=\frac{1}{\sqrt{2N}}
[\psi_1(\v{x}_1)\psi_2(\v{x}_2)-\psi_1(\v{x}_2)\psi_2(\v{x}_1)]
\ee
in non-relativistic Quantum Mechanics where $\psi_1(\v{x})$ and $\psi_1(\v{x})$
are two not necessarily orthogonal wave functions. The matrix element
\be
\la\psi|\Delta|\psi\ra
=\frac{1}{N}[\la\psi_1|\Delta|\psi_1\ra\la\psi_2|\psi_2\ra
+\la\psi_2|\Delta|\psi_2\ra\la\psi_1|\psi_1\ra
-2Re\la\psi_1|\Delta|\psi_2\ra\la\psi_2|\psi_1\ra]
\ee
shows that the kinetic energy receives a contribution from the entanglement
of the two particle states in the region where the one-particle wave functions 
overlap and this additional piece in the energy might be interpreted as an interaction 
potential
\be\label{coren}
U(\v{x}_1,\v{x}_2)=-\frac{Re\la\psi_1|\Delta|\psi_2\ra\la\psi_2|\psi_1\ra}
{\la\psi_1|\psi_1\ra\la\psi_2|\psi_2\ra-|\la\psi_1|\psi_2\ra|^2}.
\ee
All this is the well known Pauli-blocking: the anti-symmetrization
excludes certain kinematical regions leading thereby to an increase of the kinetic 
energy. This blocking is restricted to small separations since the one-particle
states must overlap. 

But this picture is static, ignoring the dynamics. The modification of the density
distribution can be viewed as a result of polarizations due to particle-anti particle
pairs whose propagation establishes non-local exchange correlation effects during some time.
The near-field vertex function obtained from the $O(\Box^{-1})$ kernel of the 
non-physical current with $\beta<0$ restricts correlations onto the light-cone.
This is what happens in QED, namely the mixing of the states of the photon and the 
electron-positron pair in a gauge and Lorentz invariant dynamics keeps the 
polarization of the Dirac-see on the mass shell $m^2=0$. But the the UV 
regime of the $O(\Box^{-2})$ kernel of the true electric current is suppressed and 
the singularity at the light-cone, ie. at vanishing invariant length 
is reduced to a finite discontinuity leaving no asymptotic particle-like states 
contributing to the current two-point function. It is remarkable that the vertex function is
actually a step function and it establishes a distance independent force between 
static world-tubes in sufficiently long time.

Notice furthermore that it is not necessary to have relativistic fermions
to find such a long-range exchange correlation. Charged relativistic bosons
display a similar dynamics, the only difference compared to the fermion case being
an overall sign in the generating functional $W[\ha]$ apart of the
modification of the current operator the external field $\ha$ is coupled to.
But both the bosonic and the fermionic currents are bilinears of the elementary 
fields and, therefore, the vacuum polarization should be qualitatively similar. 
It is not the filled states of the Dirac-see which are important but 
rather the propagation of states created by the current operator from the
vacuum. These excitations are always bosonic and thus subject to an effective
bosonic description. Non-relativistic systems, such as non-interacting fermions at 
finite density display similar effect, too.

This distance-independent force would not be observable even if one had sufficient
space resolution in the experimental device. The linear potential is a special
feature of the non-interacting Dirac-see only and becomes screened
by the electromagnetic interactions. In fact, suppose we had a matter and an anti-matter 
localized states and we started to separate the two components. Increasing the 
separation beyond the Compton wavelength of the electron costs energy comparable 
with those needed to put a virtual $e^-e^+$ pair on the mass-shell. Furthermore, 
the gradient expansion is valid for separations larger than the Compton wavelength. 
Therefore the charge of the real particles which are created by the minimal coupling 
vertex tends to screen the original charges and only localized states with non-integer 
charge remain strongly correlated.

\subsection{Localized states}\label{cpose}
The preceding discussion of the dynamics of the polarized current is restricted
to weak external sources. The involved, non-polynomial character of the generating 
functional $W^\mr{el}_0[\ha]$ makes its appearance for strong external sources. Though the
correspondence $\ha\to\hJ$ is unique but its inverse is not and the effective actions
are not uniquely defined for strong external sources. By retaining the
profile of the current only one looses important information. For instance,
a given charge density has different dynamics depending on the nature of the 
states which contribute to the charge. This is reminiscent of the small and 
large polaron problem in solid state physics where the same polarization may 
induce different dynamical responses depending on the way the polarization builds up.

For weak external sources the polarized charge is made up by small contributions
received form a number of negative energy one-particle states. The resulting
charge density can be localized but extended states contribute only. As a result,
Pauli-blocking renders the dynamics of such localized states highly correlated. 
For slightly stronger localized external field $a$ bound states are formed in the
mass gap. This case will be considered in a qualitative manner below.
For stronger localized external field the localized states may appear in the negative 
energy continuum. Such bound states appear as a violation of the convexity of the effective
action; a rearrangement of the vacuum takes place. One member of a virtual $e^+e^-$ 
pair created by the external field jumps into the potential well and the resulting 
energy is sufficient to put the other charge on mass-shell. As a result, a new, 
well defined Legendre transform of $W_0^e[\ha]$ is regained.
In other words, the effective action, $\Gamma^\mr{el}[\hJ]$, is multi-valued and may have
several consistent "sheets" for a given current. 

Let us return to the case of medium strong, localized external field which 
create localized states in the mass gap. The external fields $\hta$ and $\htab$ can be
used to place charges into the initial or final state which can be captured by these
potential centers. These charges have weak overlap with 
the filled extended states and are supposed to obey the "free dynamics" anticipated 
from classical mechanics. We start with charges of the same sign with fixed 
boundary condition where 
$W^\mr{el}[\ha]=W[\ha]^e_0-i\ln G[\ha]$, $G[\ha]$ being the valence propagator
with given initial and final points. We are interested in the possible
decoupling of the trajectories. Therefore, the initial and final location
and spin of the charges are chosen to be the same.
The external potential is chosen to be static and of the form $a_\mu=g_{\mu0}u$
where the temporal component, $u$, is the sum of potential wells which are
strong enough to create localized state in the mass gap and spread enough such that 
the characteristic size of the bound state, $\ell$, be large compared to the Compton 
wavelength of electrons, $m\ell>1$, in order to avoid pair creation. We choose $t_f-t_i\gg1/m$ 
and shall apply the non-relativistic approximation for the valence propagator.
What is crucial is that $W[\ha]^e_0$ can then be ignored in $W^\mr{el}[\ha]$ beside of the
valence propagator. Once the vacuum polarization is suppressed, the rest is a simple problem 
in non-relativistic quantum mechanics. In fact, we have at this point
\be\label{vtef}
\Gamma^\mr{el}_\mr{CFBC}[J]=-(t_f-t_i)(m+E_0)-\int_{t_i}^{t_f}dt\int d^3xu(\v{x})J^0(t,\v{x})
\ee
for $t_f-t_i\to\infty$ where $E_0$
denotes the ground state energy of electrons bound to the potential wells.
When separating the wells the charge density remains localized at the
wells and the contribution of the last term on the right hand side breaks off into a
sum of the contributions of separate wells. The correlation energy of Eq.
\eq{coren} becomes small for well separated states and therefore the contributions of
the potential wells to $E_0$ decouple.

Once the separation of localized charges is established we can consider the
issue of the dynamics of a single localized state. Let us suppose that the
current $J$ is a narrow flux tube with flux $N$ and its center follows
the world-line $x^\mu(s)$ in space-time. One interprets this as the motion
of a charge $Ne$ along the world line $x^\mu(s)$. If the value of the effective
action evaluated for this current, divided by the invariant length of the
world line $x^\mu(s)$ is independent of the time of evolution then this
ratio is the mechanical mass. Let us consider a world line without acceleration.
Then Lorentz invariance allows us to set
$\v{x}_i=\v{x}_f$. We use the same external potential as in the previous case
except that it has a single potential well only and we place a single electron in the
initial state. The generating functional for
this problem is \eq{egyel} whose value can be written as in Eq. \eq{vtef}.
The ground state energy, $E_0$, is the sum of the expectation values of the kinetic
and the potential energies. The latter cancels against the last term on the right hand
side of Eq. \eq{vtef} because $J^0(t,\v{x})$ is the particle density,
ie. the magnitude square of the ground state wave function apart of some finite time
at the beginning and at the end of the motion which give negligible
contribution in the limit  $t_i=t_f\to\infty$. We are finally left with
$\Gamma^\mr{el}[J]=-(t_f-t_i)(m+E_k)$ and the mechanical mass, the coefficient in front 
of the first term on the right hand side of Eq. \eq{ced}, can be written as
\be
M=m+\Delta m_k+\Delta m_\mr{pol}+\Delta m_\gamma,
\ee
where $\Delta m_k$ is the expectation value of the kinetic energy in the bound state
which is approximately $1/(\ell^2m)$. The second term represents the effects of the 
polarizations of the Dirac-see by the external potential coming from
$W[\ha]^e_0$. Its order of magnitude, $m\ell u$, is estimated as the product of the 
polarized charge square, assumed to be proportional to $u/m$, the ratio of the 
energy scale of the potential and the mass, the length scale of the potential well 
$\ell$ and the string tension $\sigma\approx m^2$ introduced in Section \ref{wpres}. 
Finally, $\Delta m_\gamma$ stands for the radiative corrections arising from the 
electromagnetic interaction and is $e^2/\ell$.

The mechanical mass depends on the details the elementary particles localized 
in their world-tubes. Once the coupling to the photon field is introduced the 
point-like charges develop their vacuum polarization cloud, $\ell\approx1/m$, and 
the correction to the mechanical mass becomes non-perturbative, $O(m)$, but 
uniquely defined and can be removed by renormalization of the mass.

\section{Interacting electrons and photons}\label{inteph}
After the separate discussion of the dynamics of the expectation 
value of the photon field and the electron current we embark the case of the
interacting system.

\subsection{Generating functional for connected Green functions}\label{ingfct}
The generating functional for the connected Green functions of the current and 
the photon field is
\be\label{wqed}
e^{iW[\ha,\hj]}=\int D[\hsi]D[\hsib]D[\hA]e^{i\hsib\cdot[\hG^{-1}_0
+\hs(\ha\bre-e\hA\bre)]\cdot\hsi+\frac{i}{2}\hA\cdot\hD^{-1}_0\cdot\hA
+i\hj\hs\cdot\hA+iS_{CT}}
\ee
where the minimal coupling contains the matrix
\be
\hs=\begin{pmatrix}1&0\cr0&-1\end{pmatrix}.
\ee
The perturbative renormalization is carried out by using the counterterms
$S_{CT}=S^{QED}_{CT}+S^{e\gamma}_{CT}$ where $S^{QED}_{CT}$ contains
the usual counterterm needed for the renormalization of the electron and photon
Green functions in a given order of the loop-expansion. The additional term
is chosen to be
\be
S^{e\gamma}_{CT}=-(\Delta Z_3-\alpha)e\ha\cdot\hD_0^{-1}\cdot\hA
+\hf(\Delta Z_3+\beta)\ha\cdot\hD_0^{-1}\cdot\ha
\ee
The origin of this expression becomes clear by considering the photon part
of the action, the sum of $S^{e\gamma}_{CT}$, the bare action and the photon 
self energy counterterm,
\be
S^{e\gamma}_{CT}+\frac{e^2}{2e_B^2}\hA\cdot\hD^{-1}_0\cdot\hA
=\hf\hA\cdot\hD^{-1}_0\cdot\hA+\frac{\Delta Z_3}{2}(e\hA-\ha)\cdot\hD^{-1}_0\cdot(e\hA-\ha)
+\alpha\ha\cdot\hD_0^{-1}\cdot\hA+\frac{\beta}{2}\ha\cdot\hD_0^{-1}\cdot\ha,
\ee
by using 
\be
\frac{1}{e_B^2}=\frac{1}{e^2}+\Delta Z_3.
\ee
The second term on the right hand side with $\ha=0$ is the counterterm for the photon
self energy but its dynamical origin is in the fermionic sector of the theory.
Any vector potential which couples to the electric current as the photon
field must appear in this counterterm on equal footing with $A$. This shows that
the choice $\alpha=\beta=0$ of the finite part of the counterterms is needed
to identify the current coupled to the source $a$ with the electromagnetic 
current.

The electron field can be integrated out easily in Eq. \eq{wqed}, leaving behind
\be\label{wqedfn}
e^{iW[\ha,\hj]}=\int D[\hA]e^{iW^\mr{el}[\ha-e\hs\hA]
+\frac{i}{2}\hA\cdot\hD^{-1}_0\cdot\hA+i\hj\cdot\hA+iS_{CT}}
\ee
where $W^a[\ha]$ stands for the generator functional of the connected
Green functions of the current in the non-interacting Dirac-see as given in
eqs. \eq{wnisum}, \eq{egyel} or \eq{ketel} for different boundary conditions.
The integration over the photon field can be carried out in the loop-expansion,
see Appendix \ref{loopexp} for the details. We record here the results for the
quadratic part of the generating functional in the absence of charges in the
initial state,
\be\label{ctpw}
W[\ha,\hj]=-\hf(\ha,\hj)\cdot\begin{pmatrix}
\htG&e\htG\cdot\hs\hD_0\cr e\hD_0\hs\cdot\htG&\hD\end{pmatrix}
\cdot\begin{pmatrix}\ha\cr\hj\end{pmatrix}
\ee
where
\be
\htG=\frac{1}{\htG_0^{-1}-\hat\Sigma^\mr{el}}
\ee
is the full current two-point function containing the current self-energy $\hat\Sigma^\mr{el}$
and
\be
\hD=\frac{1}{\hD_0^{-1}-\hat\Sigma^\mr{phot}}
\ee
stands for the photon propagator involving the photon self-energy $\hat\Sigma^\mr{phot}$.
The formal expressions for the self-energies up to two-loops are 
\be\label{ctpese}
\hat\Sigma^\mr{el}=e^2\hs\hD_0\hs-e^2\htG_0^{-1}\cdot\hW^{\mr{el}(4)}_D\cdot\htG_0^{-1}
\ee
and
\be\label{ctppse}
\hat\Sigma^\mr{phot}=e^2\hs(\htG_0-e^2\hW^{\mr{el}(4)}_D)\hs
\ee
where the first and second terms on the right hand sides
represent the one- and two-loop corrections with
\be
W^{\mr{el}(4)}_{Dab}=W^{\mr{el}(4)}_{abcd}(\hs i\hD_0\hs)_{cd}
\ee
being the one-loop self-energy and vertex correction to the current two-point function.

The more useful parameterizations \eq{jpar}, \eq{apar} yields the form
\be\label{rew}
\Re W[a,\ab,j,\jb]=\hf(a,j,\ab,\jb)\cdot
\begin{pmatrix}\frac{\kappa}{2}(K^\mr{tr}+K)&K\cr K^\mr{tr}&0\end{pmatrix}\cdot
\begin{pmatrix}a\cr j\cr\ab\cr\jb\end{pmatrix},
\ee
with
\be\label{inwobc}
K_\mr{OBC}=-\begin{pmatrix}\tG^\mr{ret}&e\tG^\mr{ret}\cdot D_0^\mr{ret}\cr eD_0^\mr{ret}\cdot\tG^\mr{ret}&D^\mr{ret}\end{pmatrix},
\ee
and
\be\label{inwfbc}
K_\mr{FBC}=-\begin{pmatrix}\tG^\mr{near}&e(\tG^\mr{near}\cdot D_0^\mr{near}-\Im\tG\cdot\Im D_0)\cr
e(D_0^\mr{near}\cdot\tG^\mr{near}-\Im D_0\cdot\Im\tG)&D_0\end{pmatrix},
\ee
in the absence of charges with positive energy, for open and fixed boundary conditions,
respectively.
As pointed out after Eq. \eq{eqmakap} the real part of the non-interacting
generating functional for fixed boundary condition can be obtained from
that of the open boundary condition by replacing the retarded and
advanced Green functions by the near field version. Such a change is not
sufficient for interacting system where higher loop may bring in the
product of two imaginary parts into $\Re W$. However by restricting our discussion
to distance scales beyond the Compton wavelength of the electron the creations of mass-shell 
$e^+e^-$ pairs are suppressed and $\Im\tG=0$, cf. Eq. \eq{gradetg} obtained in the framework 
of the gradient expansion. It is pointed out in Appendix \ref{raton} that this is the
only violation of the simple rule mentioned above and once the pair creation is excluded
one can again obtain easily the fixed boundary condition expressions from their
open boundary counterparts. 

The comparison of Eqs. \eq{wrepko} and \eq{wrepkf}
reveals that the imaginary part of the propagator undergoes more substantial changes 
then the real part when the boundary condition is modified.

\subsection{Legendre transformation}
We introduce a number of effective actions. The simplest effective action arises from 
the Legendre transformation of the variables $a^\pm$ and $j^\pm$,
\be
\Gamma[\hJ,\hA]=W[\ha,\hj]-\ha\cdot\hJ-\hj\cdot\hA
\ee
where
\be
\hJ=\fd{W}{\ha}
\ee
and
\be
\hA=\fd{W}{\hj}.
\ee
The inverse of the block-matrix needed for the change of variables in the generating 
functional \eq{ctpw} can be read off from the result of Eq. \eq{invbm} yielding
\be\label{pmsefa}
\Gamma[\hJ,\hA]=\Gamma^\mr{mech}[\hJ]+\hf\hA\cdot\hD^{-1}_0\cdot\hA-e\hA\hs\cdot\hJ
\ee
where the first term is responsible for the dynamics of charges moving within the 
interacting Dirac-see and is given by
\be
\Gamma^\mr{mech}[\hJ]=\hf\hJ\cdot(\htGi_0-\htG_0^{-1}\cdot\hW^{\mr{el}(4)}_D\cdot\htG_0^{-1})\cdot\hJ
\ee
on the two-loop level. The equations of motions are
\be
\fd{\Gamma[\hJ,\hA]}{\hJ}=-\ha
\ee
and
\be
\fd{\Gamma[\hJ,\hA]}{\hA}=-\hj.
\ee

What is remarkable in the expression \eq{pmsefa} is that the electromagnetic interactions
are represented without loop-corrections, ie. by the inverse of the free photon 
propagator and the minimal coupling without a form factor. This is 
because the introduction of separate variables for the current and the photon field formally
removes the one-loop contributions from the action. In fact, the elimination of one of the fields
by its equation of motion reintroduces the one-loop pieces in form of an action at a 
distance. In fact, eliminating $\hJ$ or $\hA$ by their equations of 
motion generates the action at a distance effective actions
\be\label{aadha}
\Gamma[\hA]=\hf\hA\cdot\hD^{-1}\cdot\hA
\ee
or
\be\label{aadhj}
\Gamma[\hJ]=\hf\hJ\cdot\htGi\cdot\hJ
\ee
for $\hj=\ha=0$, involving the full propagators.

Although the effective action \eq{pmsefa} shows obvious similarities with the action of
classical electrodynamics it is not satisfactory. In fact, though either
fields of the type $+$ or $-$ can be used to read off measured expectation 
values but we have both of them in the effective action coupled to each other.
The physical expectation values are easier to read off when the Legendre transformation
is performed on the variables introduced by Eqs. \eq{jpar} and \eq{apar}. For this end
we shall use the effective action
\be\label{effaaj}
\Gamma[J,J^\mr{adv},A,A^\mr{adv}]=\Re W[a,\ab,j,\jb]-J\cdot a-J^\mr{adv}\cdot\ab-A\cdot j-A^\mr{adv}\cdot\jb
\ee
with the new variables
\be
J=\fd{\Re W[\ha,\hj]}{a},~~~~~~J^\mr{adv}=\fd{\Re W[\ha,\hj]}{\ab},~~~~~~
A=\fd{\Re W[\ha,\hj]}{j},~~~~~~A^\mr{adv}=\fd{\Re W[\ha,\hj]}{\jb}.
\ee
The inverse Legendre transformation is based on the relations
\be\label{eqmj}
a=-\fd{\Gamma[J,J^\mr{adv},A,A^\mr{adv}]}{J},~~~~~~\ab=-\fd{\Gamma[J,J^\mr{adv},A,A^\mr{adv}]}{J^\mr{adv}}~~~~~~
\ee
and
\be\label{eqma}
j=-\fd{\Gamma[J,J^\mr{adv},A,A^\mr{adv}]}{A},~~~~~~\jb=-\fd{\Gamma[J,J^\mr{adv},A,A^\mr{adv}]}{A^\mr{adv}}.
\ee
The block-matrix matrix of Eq. \eq{inwobc} yields, after some straightforward calculation
outlined in Appendix \ref{efff}, the effective action
\be\label{efafvo}
\Gamma[J,J^\mr{adv},A,A^\mr{adv}]=-\hf(J,A,J^\mr{adv},A^\mr{adv})\cdot
\begin{pmatrix}0&K^{\mr{tr}-1}\cr K^{-1}&-\frac{\kappa}{2}(K^{\mr{tr}-1}+K^{-1})\end{pmatrix}
\cdot\begin{pmatrix}J\cr A\cr J^\mr{adv}\cr A^\mr{adv}\end{pmatrix},
\ee
where
\be\label{kinvo}
K^{-1}=-\begin{pmatrix}\tG^{\mr{ret}-1}_0+e^2\tG^{\mr{ret}-1}_0\cdot W^{\mr{el}(4)\mr{ret}}_D\cdot\tG^{\mr{ret}-1}_0&-e\cr-e&D^{\mr{ret}-1}_0\end{pmatrix}
\ee
for open boundary condition. The retarded part 
$W^{\mr{el}(4)\mr{ret}}_D=\Re W^{\mr{el}(4)++}_D+\Re W^{\mr{el}(4)-+}_D$ 
is defined in the same manner as for the free block-propagator in Eq. \eq{photpr},
cf. Appendix \ref{raton}.
The equations of motion \eq{eqmoff} give $J^\mr{adv}=A^\mr{adv}=0$ for vanishing
off-shell sources, $a=j=0$, as expected, and we find the equations of motion 
\bea\label{emff}
J&=&\frac{1}{1+e^2W^{\mr{el}(4)r}_D\cdot\tG^{\mr{ret}-1}_0}\cdot\tG^\mr{ret}_0\cdot(eA-\ab),\nn
A&=&D^\mr{ret}_0\cdot(eJ-\jb)
\eea
for the physical fields. The one-loop contributions are again absent
and the two-loop term represents the electromagnetic interaction in the Dirac-see
in the first line of Eqs. \eq{emff} as in the effective action \eq{pmsefa}. According to the 
second line of Eqs. \eq{emff} both the external and the dynamical
currents,  $\jb$ and $J$ induce retarded potentials.

In the case of fixed boundary condition we replace the retarded and advanced
propagators by the near field version in the absence of creation of mass shell $e^+e^-$ pairs
and use
\be\label{kinvf}
K^{-1}=-\begin{pmatrix}\tG^{\mr{near}-1}_0+e^2\tG^{\mr{near}-1}_0\cdot W^{\mr{el}(4)\mr{near}}_D\cdot\tG^{\mr{near}-1}_0&-e\cr-e&D^{\mr{near}-1}_0\end{pmatrix}
\ee
in Eq. \eq{efafvo}. The equations of motion become
\bea\label{emfff}
\tG^\mr{near}_0\ab&=&-(1+e^2W^{\mr{el}(4)\mr{near}}_D\cdot\tG^{\mr{near}-1}_0)\cdot J+e\tG^\mr{near}_0\cdot A,\nn
D_0^\mr{near}\jb&=&eD_0^\mr{near}J-A.
\eea

The simplest effective action includes physical fields only and it is defined as
\be\label{effaj}
\Gamma[J,A]=\Re W[a,\ab,j,\jb]-J\cdot a-A\cdot j
\ee
with the new variables
\be
J=\fd{\Re W[\ha,\hj]}{a},~~~~~~A=\fd{\Re W[\ha,\hj]}{j},
\ee
and the source $\ab$ and $\jb$ treated as passive parameters in the Legendre transformation.
Notice that the important block in the generating functional \eq{rew} we have to invert
to find the equations of motion for the physical field, the upper left
one, is the same for both boundary conditions. The dependence on the boundary condition appears
through the linear pieces of the effective action only. The simple steps shown in 
Appendix \ref{eftf} lead to the effective action
\bea\label{efaffvo}
\kappa\Gamma[J,A]&=&\Gamma^\mr{mech}[J]+\hf A\cdot D^{\mr{near}-1}_0\cdot A-eA\cdot J-A\cdot J^\mr{ext},\nn
\Gamma^\mr{mech}[\hJ]&=&\hf J\cdot(\tG^{\mr{near}-1}_0+\tG_0^{\mr{near}-1}\cdot W^{\mr{el}(4)\mr{near}}_D\cdot\tG_0^{\mr{near}-1})\cdot J-J\cdot A^\mr{ext}
\eea
where the external sources
\bea
J^\mr{ext}&=&D^{\mr{near}-1}_0\cdot W_j-eW_a\nn
A^\mr{ext}&=&(\tG^{\mr{near}-1}_0+\tG_0^{\mr{near}-1}\cdot W^{\mr{el}(4)\mr{near}}_D\cdot\tG_0^{\mr{near}-1})\cdot W_a-eW_j
\eea
are given in terms of the boundary condition dependent terms $W_a$, $W_j$ shown in Eqs. \eq{wtfa} and
\eq{wtfj}. The corresponding equations of motion for the field $J$ is
\be\label{emffj}
J=-\tG^{\stackrel{\mr{ret}}{\mr{near}}}(\ab+eD_0^{\stackrel{\mr{ret}}{\mr{near}}}\cdot\jb)
+\frac{e}{\tG^{\mr{near}-1}_0+\tG^{\mr{near}-1}_0\cdot W^{(4)n}_D\cdot\tG^{\mr{near}-1}_0}
[A+D^{\stackrel{\mr{ret}}{\mr{near}}}\cdot\jb+eD^{\stackrel{\mr{ret}}{\mr{near}}}_0\cdot\tG^{\stackrel{\mr{ret}}{\mr{near}}}\cdot\ab)]
\ee
where the upper (lower) indices correspond to open (fixed) boundary condition.
The equation of motion for the electromagnetic field is
\be\label{emffao}
A=eD_0^\mr{near}\cdot J-D^\mr{ret}_0\cdot\jb+\frac{e}{2}D^\mr{far}_0\cdot\tG^\mr{ret}\cdot(\ab+eD_0^\mr{ret}\cdot\jb)
+(e^4D_0^\mr{ret}\cdot W^{\mr{el}(4)\mr{ret}}_D\cdot D_0^\mr{ret}-e^4D_0^\mr{ret}\cdot\tG_0^\mr{ret}\cdot D_0^\mr{ret}\cdot\tG_0^\mr{ret}\cdot D_0^\mr{ret})\cdot\jb
\ee
for open boundary condition and
\be\label{emffaf}
A=eD_0^\mr{near}\cdot J-[D_0^\mr{near}-e^4D_0^\mr{near}\cdot W^{\mr{el}(4)\mr{near}}_D\cdot D_0^\mr{near}
+e^4D_0^\mr{near}\cdot\tG_0^\mr{near}\cdot D_0^\mr{near}\cdot\tG_0^\mr{near}\cdot D_0^\mr{near}]\cdot\jb
\ee
for closed boundary condition.

Notice the difference between Eqs. \eq{emff} and \eq{emffj}-\eq{emffao}, the
equations of motion for the physical fields derived from the actions 
$\Gamma[J,J^\mr{adv},A,A^\mr{adv}]$ and $\Gamma[J,A]$ with open boundary conditions.
We shall come back to this apparent paradox of having different
equations for the same quantity in Section \ref{qrg}.

\section{Microscopical classical field theory}\label{miclth}
It is important to distinguish two length scales when constructing classical field 
theories from quantum theory.
The shorter one, $a=2\pi\Lambda^{-1}$, is the UV cutoff of the underlying Quantum Field Theory.
The other is the quantum-classical crossover length scale, $\xi_\mr{cr}$, which
is a rough order of magnitude estimate since the scale of the actual crossover
depends on the environment. Classical field theories derived above for the 
expectation values of local operators provide information about the dynamics with 
a space-time resolution limited by $a$ only. Structures seen at scales
$a<\ell<\xi_\mr{cr}$ characterize the microscopic quantum dynamics. Our classical 
field theory can be viewed as a time dependent generalization of the density
functional theory \cite{hoko,kosa,lieb,nale,fukuda,fukuk,raja,rajak,valiev,posa}
for relevant observables such as the electromagnetic 
current \cite{cfto,cfte} and the electromagnetic field. The current
and electromagnetic field profiles, given by the equations of motion, help us to 
perform a partial resummation of the perturbation expansion. The
study of the expectation values at scales $\ell\approx\xi_\mr{cr}$ should
give us some new insight into the quantum-classical transition because 
on the one hand, the description includes a large number of degrees of freedom
needed to cope with classical objects and on the other hand, it is based on
expectation values which are the relevant quantities at both sides of the transition.
Notice that this is not the case when the usual, scattering amplitude based 
formalism of Quantum Field Theory is used because it is restricted to pure states 
and the decoherence can not even be formulated properly. Finally, we see macroscopic 
physics for $\xi_\mr{cr}<\ell$. 

Three aspects of microscopic classical theory will be mentioned in this Section.
One is the decoherence and another, induced by it, the transfer of the time arrow
by the environment. The third issue is the treatment of the polarization
effects of the Dirac-see.

\subsection{Decoherence of non-relativistic particles}\label{decohnrc}
Let us consider, as a simple but non-trivial problem, a system of $N$ non-relativistic charges
interacting with the photon field. The dynamics of charges is 
characterized by the action $S_c[\v{x}]$ and the coupling to the electromagnetic
field is realized via the current 
\be
j[\v{x}]_{\mu x}=\sum_{n=1}^N\delta_{x^0,t}\delta_{\v{x},\v{x}^{(n)}(t)}(1,\dot{\v{x}}^{(n)}(t))
\ee
where $\v{x}^{(n)}(t)$ denotes the trajectory of the $n$-th charge. We consider
the functional integral
\be
Z=\int D[\hA]D[\hat{\v{x}}]e^{\frac{i}{2}\hA\cdot\hD^{-1}_0\cdot\hA
+iS[\hat{\v{x}}]-ie\hat j[\v{x}]\hs\cdot\hA}
\ee
where $D[\hat{\v{x}}]$ denotes the integral over the trajectories of all charges, 
$\hat{\v{x}}=(\v{x}^+,\v{x}^-)$ stands for the particle trajectory,
$S[\hat{\v{x}}]=S[\v{x}^+]-S^*[\v{x}^-]$ and
the current $\hat j[\v{x}]=(j[\v{x}^+],j[\v{x}^-])$ couples to the photon field.

The photons can be eliminated in the usual way 
and the result is the action at a distance form of the electromagnetic interactions
\be
Z=\int D[\hat{\v{x}}]e^{iS_\mr{eff}[\hat{\v{x}}]},
\ee
with
\be
S_\mr{eff}[\hat{\v{x}}]=S[\hat{\v{x}}]+W^\mr{phot}[-ej[\v{x}^+],ej[\v{x}^-]]
\ee
where the last term on the right hand side is the influence functional \cite{feve}.
The form \eq{phw} for the generating functional $W_\mr{OBC}^\mr{phot}[\hj]$ gives
\be\label{nrchwa}
S_\mr{OBC}[\hat{\v{x}}]=S[\hat{\v{x}}]
-\frac{e^2}{2}j^\sigma[\v{x}]\cdot D_0^\mr{ret}\cdot\jb^\sigma[\v{x}]
-\frac{e^2}{2}\jb^\sigma[\v{x}]\cdot D_0^\mr{adv}\cdot j^\sigma[\v{x}]
-\frac{e^2}{2}j^\sigma[\v{x}](\kappa D_0^\mr{near}+i\Im D_0)j^\sigma[\v{x}]
\ee
for open boundary condition where the parameterization \eq{jpar} of the
current coupled by the matrix $\hs$ gives $j^\sigma=j^+-j^-$ and 
$\jb^\sigma=j^+(1-\kappa)/2+j^-(1+\kappa)/2$. 
The relations \eq{phreg}, \eq{phimg} allow us to rewrite the exponent of the 
integrand for open boundary condition in terms of the effective actions as
\be\label{atef}
S_\mr{OBC}[\hat{\v{x}}]=S[\hat{\v{x}}]+\Gamma_\mr{OBC}^\mr{phot}[A,A^\mr{adv}]
+i\Gamma_\mr{OBC}^{\mr{phot~im}}[A^\mr{im}]
-ej^\sigma[\v{x}]\cdot(A+iA^\mr{im})-e\jb^\sigma[\v{x}]\cdot A^\mr{adv},
\ee
where the fields $A$, $A^\mr{adv}$ and $A^\mr{im}$ are the solutions of the classical equations 
of motion \eq{ivtrre}, \eq{ivtrrei}. The classical field theory based on the effective actions
$\Gamma_\mr{OBC}^\mr{phot}[A,A^\mr{adv}]$ and $\Im\Gamma_\mr{OBC}^\mr{phot}[A^\mr{im}]$ allows us to obtain the
Wilsonian effective action for the charges in terms of the expectation value
of the electromagnetic field, $A$, and other non-physical expectation values,
$A^\mr{adv}$ and $A^\mr{im}$, induced by the trajectories $\hat{\v{x}}$. 

The imaginary part of the influence functional \eq{nrchwa} is due to the continuous spectrum. 
It results from the construction of the reduced density matrix, the measurement of the 
unit operator, and represents a suppression
\be
s=e^{\frac{e^2}{2}(j[\v{x}^+]-j[\v{x}^-])\cdot\Im D_0\cdot(j[\v{x}^+]-j[\v{x}^-])}
\ee
of the off-diagonal fluctuations in the density matrix $\rho(\v{x}^+,\v{x}^-)$.
The imaginary part of the two current point function receive contributions form mass shell intermediate states,
real photons, in this case. Non-accelerating charges do not emit photons;
it is easy to check that $s=1$ in this case.

Let us now consider a charge moving along a the trajectory $\v{x}(t)=\v{r}\cos\omega t$, the
corresponding current being $j^\mu_{t,\v{x}}=e\delta(\v{x}-\v{x}(t))(1,\dot{\v{x}}(t))$.
We shall calculate the imaginary part of the influence functional for 
$j^\pm_{t,\v{x}}=j_{t,\v{x}\pm\v{R}/2}$, $r=|\v{r}|\ll|\v{R}|=R$ in Fourier space where
\be
j^\mu_q=\left(eB(q^0,\omega,\v{q}\v{r}),
\v{r}\frac{ie\omega}{2}[B(q^0-\omega,\omega,\v{q}\cdot\v{r})-B(q^0+\omega,\omega,\v{q}\cdot\v{r})]\right)
\ee
with
\be\label{bapr}
B(q^0,\omega,a)=\int_te^{-iq^0t+ia\cos\omega t}=\delta(q^0)+ia\delta(q^0-\omega)+ia\delta(q^0+\omega)+O(a^2),
\ee
and the suppression factor can be written as
\be
s=e^{-2\pi\int_q\delta(q^2)|j_q|^2\sin^2\frac{\v{q}\cdot\v{R}}{2}}.
\ee
The diffraction integral in the exponent shows clearly the origin of the decoherence,
the interference between the photons emitted by $j^+$ and $j^-$. The power series on
the right hand side of Eq. \eq{bapr} corresponds to the multipole expansion and
we keep the dipole field only. Straightforward calculation gives in the limit $t_f-t_i\to\infty$
\be
s_\parallel=e^{4e^2(t_f-t_i)r^2\left[\left(\frac{\omega^2}{R}-\frac{2}{R^3}\right)\sin\omega R
+\frac{2\omega}{R^2}\cos\omega R-\frac{\omega^3}{3}\right]}
\ee
when $\v{R}$ and $\v{r}$ are parallel. For $\v{R}\cdot\v{r}=0$ one finds 
\be
s_\perp=e^{4e^2(t_f-t_i)r^2\left[\frac{1}{R^3}\sin\omega r-\frac{\omega}{R^2}\cos\omega r-\frac{\omega^3}{3}\right]}.
\ee
The suppression factor interpolates smoothly between
$s_\parallel=s_\perp=1$ for $R=0$ and 
\be\label{chdec}
s_\parallel=s_\perp=e^{-\frac{4}{3}e^2(t_f-t_i)r^2\omega^3}
\ee
for $R\approx\infty$.

\subsection{Radiation time arrow}
The Lagrangian of QED is formally invariant with respect to the inversion
of the direction of time, nevertheless our daily experience confirms that the
currents generate retarded electromagnetic fields and the time reversal
invariance is lost \cite{zeh}. This symmetry breaking must come either from the environment,
represented here as external sources which couple to the current and the 
electromagnetic field or from the way the dynamical problem is posed ie.
the boundary conditions in time.

The CTP formalism is well suited for the investigation of this problem
because of the duplication of the degrees of freedom. In the decomposition
\eq{fieldecomp} of the field variable $\phi$ picks up the diagonal
quantum fluctuations in the density matrix which appear in the expectation 
values of the canonical coordinates and $\phi^\mr{adv}$ collects off-diagonal fluctuations,
displayed by the expectation value of the canonical momentum variables. 

But there is another way of looking at this separation: from the point of view of the
time arrow. Let us return to the remark made in Section \ref{exalqft} that the time 
arrow is different for the coordinates $\phi^\pm$ appearing in the time ordered
and anti-time ordered part in Eq. \eq{gfhr}. 
Complications arise from the fact that matrix elements of time dependent 
operators include both advanced and retarded effects. For instance, the Green 
function $\la0|T[\phi(x)\phi(y)]|0\ra$ displays both advanced and retarded parts.
The discussion of the linear response formulae in Section \ref{exalqft} makes it 
clear that any reflecting final boundary conditition induces advanced fields. 
Can one separate the retarded and advanced effects?
The answer is trivial and affirmative for non-interacting particles and the decomposition 
\eq{fieldecomp} gives $\phi$ and $\phi^\mr{adv}$ as the combinations of the canonical coordinates 
which receive retarded or advanced effects of the external sources, respectively, for open boundary
condition, according to Eqs. \eq{eqmaakap} and \eq{eqmakap}. The retarded nature of the
physical field, $\phi$, can be understood by noting that the in-field, observed
at the initial time, is vanishing due to the choice $|\Psi_i\ra=|0\ra$. The separation
of the advanced and retarded effects is trivial because it can be achieved by
inspecting the initial (in) or final (out) field in the absence of vertices.

Interactions mix advanced and retarded effects because the non-linear pieces of the
effective action in the fields are time reversal invariant. Consider for instance 
the equations of motion arising from the effective action $\Gamma[J,A]$ of Eq. \eq{efaffvo}.
The time reversal invariance is broken by the environmental variables $\ab$ and $\jb$ 
which appear only in the terms linear in the fields, the part of the effective
action which depends on the boundary conditions in time. The relation between the 
expectation values $J$ and $A$, given by the boundary condition independent quadratic 
part of the effective action, involves the time reversal invariant near field 
propagator. The local expectation values appear to be insensitive to the symmetry 
breaking effects of the boundary condition. Our conjecture is that closed quantum 
systems maintain time reversal invariance for any boundary condition in time.

Such a claim makes the coupling to the classical environment responsible for the choice
of the time arrow. But the environment characterized by the external sources $\ab$ and 
$\jb$ obey quantum laws, as well, suggesting that it is the classical limit for the
environment which triggers the breakdown of the time reversal invariance. In fact, the
decoherence which is supposed to be the hallmark of the quantum-classical
crossover suppresses advanced fields without influencing 
retarded combinations, according to Eq. \eq{nrchwa}. Notice the double
role the imaginary part of the influence functional plays in Eqs. \eq{nrchwa}-\eq{atef}.
It sets not only the direction of time of the decoherence but determines the 
proportion of retarded and advanced parts in the charge propagator, too.

The equations of motion of the effective actions $\Gamma[J,J^\mr{adv},A,A^\mr{adv}]$ support the view
that the retarded and advanced propagators appear in the equations of motion
due to decoherence. On the first sight it is rather confusing that the quadratic part 
of the effective action in the fields involve the retarded and advanced Green functions 
when open boundary condition is used, in contrast to $\Gamma[J,A]$. How can different 
equations of motion occur at all for the same expectation values? The only 
difference between the equations of motion for $J$ and $A$, arising from the
effective actions $\Gamma[J,J^\mr{adv},A,A^\mr{adv}]$ and $\Gamma[J,A]$ is that different 
quantities are kept fixed during the variations. In fact, let us
construct a variation of $A$ by varying the external sources $a$ and $j$ 
for fixed $\ab$ and $\jb$ in such a manner that $J$ remains unchanged. Such a 
variation gives rise to the equation of motion for $A$, Eq. \eq{emffao}, of the 
effective action $\Gamma[J,A]$. In the case of the effective action
$\Gamma[J,J^\mr{adv},A,A^\mr{adv}]$ all four sources are varied by keeping all field but $A$ fixed. 
The lesson is that the time arrow appeared in the equations of motion relating
measured expectation values because the variations respected the condition
$J^\mr{adv}=A^\mr{adv}=0$, the vanishing of the off-diagonal fluctuations. This is exactly what
decoherence is for.

Can the decoherence solve the arrow of time problem? Not, it makes one step only
in reducing it to the environment. When the photons are eliminated in Section
\ref{decohnrc} then their boundary condition in time fixes the sign of $\Im D_0$
and consequently the arrow of time for the charges. What is remarkable is that this 
transfer of symmetry breaking becomes enhanced as we enter the macroscopic domain 
by decreasing the cutoff. This suggests that a better understanding of the arrow 
of time problem can be obtained by the application of the renormalization group method to this problem.

\subsection{Vacuum polarization}\label{vacpol}
In classical electrodynamics the polarizability of a medium is taken into account 
in the well known, simple manner supposing the higher multipole moments are negligible.
It is interesting to see what kind of approximation corresponds to this procedure
when the polarizability of the Dirac-see is considered.

The effective action for the electromagnetic field,
\be\label{efaffvog}
\Gamma[A]=\hf A\cdot(D^{\mr{near}-1}_0-e^2\tG_0^{\mr{near}-1}+e^2\tG_0^{\mr{near}-1}\cdot W^{\mr{el}(4)\mr{near}}_D\cdot\tG_0^{\mr{near}-1})
\cdot A-eA\cdot\Je+O(A^3)+O(e^4),
\ee
obtained by eliminating the current $J$ by its equation of motion and reintroducing 
the elementary charge $e$ in the minimal coupling in Eq. \eq{efaffvo}. We ignore the 
gauge fixing term because current conservation decouples the gauge sector
and write $\Gamma[A]=\tilde\Gamma[F]$, $F_{\mu\nu}=\partial_\mu A_\nu-\partial_\nu A_\mu$
due to gauge invariance. A conserved current is defined as
\be
j=-\frac1e\fd{\Gamma[A]}{A}_{|A=0}
=\frac1e\left(\partial^\nu\fd{\tilde\Gamma[F]}{F^{\mu\nu}}-\partial^\nu\fd{\tilde\Gamma[F]}{F^{\nu\mu}}\right)_{|F=0}.
\ee
The minimal coupling is constructed by the introduction of the tensor
$H_{\mu\nu}=\partial_\mu j_\nu-\partial_\nu j_\mu$ for the current 
and writing the $O(A)$ part of the effective action as
\be
-eA\cdot j=\frac{e}{2}F_{\mu\nu}\cdot\Box^{-1}\cdot H^{\mu\nu}
\ee
where the conservation of the currents was used on the right hand side.
The field strength tensor which includes the vacuum polarization is defined as
\be
G=-2\left(\fd{\tilde\Gamma[F]}{F}-\fd{\tilde\Gamma[F]}{F}_{|F=0}\right).
\ee
To find the linearized equation of motion for the photon field it is 
sufficient to have the accuracy $O(J)$ and $O(A^2)$ for the effective action,
\be
\Gamma[A]=-\frac14F\cdot G-eA\cdot j+O(A^3).
\ee

To find a more detailed expression for the field strength tensor $G$ return to
the effective action $\Gamma[A]$ and assume that its $O(A^2)$ part is translational
invariant. This latter property allows us to write
\be
\hf A\cdot T\cdot\Gamma^{(2)}\cdot T\cdot A=-\frac18F_{\mu\nu}\cdot\frac{1}{\Box}[\Gamma^{(2)},F]_+^{\mu\nu}
-\frac18F_{\mu\rho}\cdot L^\rho_{\ \ \nu}\cdot\frac{1}{\Box}[\Gamma^{(2)},F]_-^{\mu\nu}
+\frac18F_{\rho\mu}\cdot L^\rho_{\ \ \nu}\cdot\frac{1}{\Box}[\Gamma^{(2)},F]_-^{\mu\nu}
\ee
where
\be
([\Gamma,F]_\pm)_x^{\mu\nu}=\int_y[\Gamma^{\mu\rho}_{x-y}F_{\rho\ y}^{\ \ \nu}
\pm F^{\mu\rho}_y\Gamma_{\rho\ x-y}^{\ \ \nu}]
\ee
the integration over the space-time coordinate $y$ being shown explicitly.
We write $\Gamma^{(2)}=-D^{\mr{near}-1}_0+\Sigma$, cf. Eq. \eq{efaffvog},
$\Sigma$ being the photon self energy, and find
\be
G=F+\hf[\Sigma,\Box^{-1}F]_+-\hf L\cdot[\Sigma,\Box^{-1}F]_-
+\hf(L\cdot[\Sigma,\Box^{-1}F]_-)^\mr{tr}
\ee
where the superscript tr denotes the exchange of the two space-time indices in the 
last term and renders $G$ an antisymmetric tensor.

It is instructive to obtain the field strength tensor in three-dimensional notations
when $A_\mu=(\phi,-\v{A})$, $\v{E}=-\partial_0\v{A}-\nabla\phi$,
$\v{B}=\nabla\times\v{A}$ and the parameterization
\be\label{phsepar}
\Sigma_{\mu\nu}=\begin{pmatrix}\Sigma_0&\v{\Sigma}\cr\v{\Sigma}&\Pi\end{pmatrix}
\ee
yield
\bea\label{poltgen}
G^{\mu\nu}&=&F+\hf\Box^{-1}\begin{pmatrix}0&(\Pi-\Sigma_0)\v{E}-\v{\Sigma}\times\v{B}\cr
(\Sigma_0-\Pi)\v{E}-\v{\Sigma}\times\v{B}&
\v{\Sigma}\otimes\v{E}-\v{E}\otimes\v{\Sigma}-[\Pi,\Phi[\v{B}]]_+\end{pmatrix}\nn
&&+\hf\Box^{-2}\begin{pmatrix}0&\v{\partial}(2\partial_0\v{\Sigma}\cdot\v{E}+\v{\partial}\cdot\v{v})
-\partial_0(\partial_0\v{u}+\v{\partial}w)\cr
-\v{\partial}(2\partial_0\v{\Sigma}\cdot\v{E}+\v{\partial}\cdot\v{v})
+\partial_0(\partial_0\v{u}+\v{\partial}w)&
\partial_0\partial^ku^j-\partial_0\partial^ju^k
+\partial^\ell\partial^kw^{\ell j}-\partial^\ell\partial^jw^{\ell k}\end{pmatrix}
\eea
with
\bea
\v{u}&=&-(\Pi+\Sigma_0)\v{E}-\v{\Sigma}\times\v{B}\nn
\v{v}&=&-(\Pi+\Sigma_0)\v{E}+\v{\Sigma}\times\v{B}\nn
w&=&\v{\Sigma}\otimes\v{E}+\v{E}\otimes\v{\Sigma}-[\Pi,\Phi[\v{B}]]_-
\eea
$\Phi[\v{B}]_{jk}=\epsilon_{jk\ell}B_\ell$ being the spacelike part of the bare
field strength tensor $F$. 

The motion of the charges in the Dirac see mixes the electric and the magnetic fields
in the vacuum polarization if translational invariance is assumed only. The rather
involved form of the polarization shown in Eq. \eq{poltgen} results from the
Schwinger-Dyson resummation of the one-loop photon self energy parameterized as \eq{phsepar}.
It is interesting to note that similar resummation has already been used in
classical electrodynamics in arriving at the Clausius-Mossotti formule \cite{jackson}.

The polarization simplifies considerably when rotational invariance, 
$\Pi_{jk}=\Pi\delta_{jk}$, $\v{\Sigma}=0$ imposed. Expression \eq{poltgen} results in
\be
\v{D}=\v{E}+\Box^{-1}\left(\Sigma_0-\hf\Sigma_\mr{nr}\cdot[1+\Box^{-1}
(\partial^2_0-\v{\partial}\otimes\v{\partial})]\right)\v{E}
\ee
and 
\be
\v{H}=\v{B}+\Box^{-1}\left[(\Sigma_0-\Sigma_\mr{nr})\v{B}+\hf\Sigma_\mr{nr}\Box^{-1}\partial_0\v{\partial}\times\v{E}\right],
\ee
respectively, where $\v{D}$ and $\v{H}$ are the electric and the magnetic fields 
contained by the field strength tensor $G$, and $\Sigma_\mr{nr}=\Pi+\Sigma_0$.

Further simplification is achieved by imposing Lorentz invariance, $\Sigma_\mr{nr}=0$, 
because the electric and magnetic sectors no longer mix. The dielectric tensor 
$\epsilon$ and magnetic permeability tensor $\mu$ defined by the relations 
$\v{D}=\epsilon\cdot\v{E}$ and $\v{B}=\mu\cdot\v{H}$, respectively, are 
\be
\epsilon=\mu^{-1}=\openone+\Box^{-1}\Sigma_0,
\ee
reproducing well known one-loop result \cite{uehl}.
Notice that the operator $\Box^{-1}$ on the right hand side is the result of
gauge invariance, the relation between the gauge field and the field strength tensor.

The $n$-th multipole moments of the vacuum polarization are constructed by
projecting the $\ell=n$ angular momentum components out of the current $n$-point 
Green function. The dipole contribution corresponds to the retaining of the
$O(a^2)$ part of the generating functional $W[a,j]$. The order of the retained
dependence on $j$ gives the accuracy in terms of the powers of $e$,
according to Eq. \eq{wjale}. The effective action 
formalism where the Legendre transformation performs the inversion of the 
quadratic part, the resummation of the Schwinger-Dyson equation, yields the 
relation between the applied and the polarized fields. 
The total field strength \eq{poltgen} is based on the dipole contribution to the 
polarization taken into account in the same order in $e$ as the photon self energy \eq{phsepar}.
Higher multipole moments introduce higher order terms in the fields in the effective
action and lead to non-linear polarization effects. Note that the $n$-th
order multipole contributions do not appear before the order $n$ of the
expansion in $e$.

We have traced the effects of vacuum polarization in the real part of the 
influence functional for the photon field which give the quantum corrections
to the classical Maxwell equations. The imaginary part of this functional
eg. $\Im\epsilon$ in a relativistically invariant theory
plays important role, as well, by introducing decoherence and setting the 
time arrow for the electromagnetic field.

\section{Quantum renormalization group}\label{qrg}
The problem of the quantum-classical transition can be studied by adjusting
the resolution of the space-time averages which is usually realized within the
framework of the renormalization group method. This method was developed in its full 
generalities first when the Kadanoff-Wilson blocking was applied to the description
of critical phenomena in classical Statistical Physics \cite{grw}. 
The application of blocking in Quantum Field Theory followed immediately and agreement
with the already known multiplicative renormalization group scheme was established.

However, there is an essential difference between the successive elimination of degrees 
of freedom in a classical and a quantum system which was ignored. In classical statistical 
physics the form of the partition function can be maintained during the blocking but 
the vacuum-to-vacuum transition amplitude of a quantum system is changed 
in a fundamental manner when particle modes are eliminated. In fact, the transition 
amplitude between two states corresponds to a process occuring for pure states while 
in Quantum Mechanics the elimination of a degree of freedom generates mixed states
to be described by a density matrix. This goes beyond the usual
formalism based on "in-out" type vacuum-to-vacuum transition amplitudes.

Let us follow the blocking procedure on the path integral \eq{piwdefe} where 
some high energy particle modes, $p>\Lambda$, are integrated over. As long as the
the initial and final states are pure the density matrices factorize in the
variables $\phi^\pm$ and the successive integration over particle modes of the
fields $\phi^\pm$ preserves this factorization. Notice that such a blocking corresponds
to the elimination of the particle modes which possess given initial and
final states coded by the density matrices. Let us suppose, for example, that the
occupation numbers for the high energy modes to be eliminated are fixed in the 
initial and final states. Then the resulting path integral displays a dynamics for 
the low energy particles which depends on these occupation numbers. 

This is not what is meant by elimination of a degrees of freedom in Quantum Mechanics. 
The measurement of the identity operator, the calculation of the
trace of the density matrix in the factor space of the degrees of freedom to 
be eliminated, destroys all information about the initial and final states.
The information about the final state is destroyed by the trace operation, the
dependence on the initial state disappears due to the unitarity of the time evolution.

It is this point, the realization of such a true elimination of degrees of freedom
without retaining information about the initial and final states, where the CTP 
formalism becomes essential. We erase the final state information for the 
high energy modes to be eliminated by installing the identity operator as density matrix
in that sector of the Fock space. The identity operator, the functional
Dirac-delta for the given modes of $\phi^+-\phi^-$ couples the variables along the
two time axis and this coupling is handed over for the low energy modes. In other
words, once we perform a true trace operation on the high energy sector of the Fock space
the remaining low energy states become mixed. This mixing requires the
use of the CTP formalism for the rest of the blocking procedure.

The difference between the blocking in the vacuum-to-vacuum amplitude and in the
CTP path integral is that in the former case the trace operation is performed in 
the zero particle number sector of the high energy modes only. This represents
a small error when the the particle modes to be eliminated are at very high energy
compared with the observational scale and perturbation expansion applies.
But in non-asymtotically free models where the high energy dynamics is
non-perturbative or in effective theories with not too high cutoff $\Lambda$ 
the weight of the particles at the cutoff scale is non-negligible and the difference between these two blocking
becomes more important. It is the quantum renormalization group procedure, based on the
CTP formalism, what should be followed in these cases.

There is no difficulty in extending the functional renormalization group scheme 
\cite{wetteq,morriseq} to the CTP formalism \cite{rge,rgk} but the doubling
of the degrees of freedom renders the resulting renormalized trajectory formal.
A more natural formalism, based on the measurable expectation values
and the degrees of freedom controlling decoherence, is outlined in
Appendix \ref{frgctp}. But instead of embarking a detailed study 
of this evolution equation we shall be satisfied by a few qualitative remarks
about the emergence of classical physics. The 
renormalized trajectory of realistic models passes at several fixed points and may 
experience different scaling laws separated by crossover regimes \cite{grg}. One of
the crossovers belongs to the quantum-classical transition. The difference between 
the trajectories generated by the blocking in the traditional, scattering
amplitude based formalism and 
the CTP path integral should be less important at the UV side of the quantum-classical
crossover. The decoherence which requires the density matrix is supposed to be
the hallmark of the quantum-classical crossover. Therefore one expects
that this transition and possible others further down towards the infrared 
direction are missed completely or reproduced with substantial errors by the 
"in-out" formalism which can not handle mixed states. Note that we need
a bosonized form of the theory where the excitations are handled by
bosonic fields in order to embark the problem of the quantum-classical
crossover because the expectation values of fermionic operators are trivially 
vanishing.

The decoherence suppresses the fluctuations of the field $\phi^\mr{adv}$.
But it is just the first moment, the expectation value of this kind of fields 
which is suppressed by construction in deriving the variational equations of motion
of the effective action $\Gamma[J,J^\mr{adv},A,A^\mr{adv}]$. This suggests that these equations
which already assume decoherence change less at the quantum-classical crossover than 
the variational equations of the effective action $\Gamma[J,A]$ which do involve 
non-vanishing advanced fields.

It is worth mentioning that the $O(A^2)$ part of the action $\Gamma[J,A]$,
given by Eq. \eq{efaffvo}, has already been suggested many years ago \cite{schwa,tetr,fok}. 
By analogy of the steps leading from Eq. \eq{pmsefa} to Eq. \eq{aadhj} we find 
for vanishing external sources the well known action at a distance
\be\label{aadhjt}
\kappa\Gamma[J]=-\hf J\cdot\tG^{\mr{near}-1}\cdot J=\Gamma^\mr{mech}[J]+\frac{e^2}{2}J\cdot D_0^\mr{near}\cdot J,
\ee
with time reversal invariant current-current interaction term. The earlier proposals 
were made on the basis of classical electrodynamics where the charges follow world 
lines and the self-interaction was excluded in order to avoid UV divergences. 
In Eq. \eq{aadhjt} self-interaction is included. Such theories have no radiation 
field, accelerating charges induce near field only. The retarded and advanced fields 
appear with the same weight and therefore the action-reaction ballance is obviously satisfied. 
The well established retarded radiation field can be recovered
by taking into account the reaction of all charges to the electromagnetic field
in a completely absorbing Universe \cite{wheeab}. As mentioned above, 
the boundary conditions in time influence the linear part of the action only. 
As a result, the $O(A^2)$ part is the same for open ended time 
evolution and for fixed boundary condition when the photons end up at the vacuum 
state in a completely absorbing Universe.
In other words, the classical argument of Ref. \cite{wheeab} applies to
open boundary condition imposed on the quantum level, as well. The time arrow and the 
Abraham-Lorentz force \cite{barut,rorl,nocorr,dirac} are generated by 
the presence of other than the accelerating charge in the system. But this argument
is valid in the microscopic domain only. As soon as the cutoff is
lowered and our action is given in the classical domain, decoherence is
implied automatically and the time arrow is generated by the retarded 
Li\`enard-Wiechert potential of the action $\Gamma[J,J^\mr{adv},A,A^\mr{adv}]$ and the Abraham-Lorentz
force is generated by the accelerating charge itself.

The renormalization group approach reveals a similarity between
spontaneous symmetry breaking and the arrow of time problem. Spontaneous
symmetry breaking appears in the renormalized trajectories as a crossover
to an infrared scaling regime which contains relevant operators of lower symmetry.
The boundary condition in time imposed on photons is passed over the charges
when photons starts to behave classically at the quantum-classical crossover.
At the final count, the arrow of time problem appears as a spontaneous 
breakdown of the time inversion symmetry. As such, it indicates that an extreme 
sensitivity develops in the system among the infrared modes to choose a time arrow
from "outside". The identity of thermodynamical arrows formed in different
domains and the importance of the cosmological master arrow can be explained by such a mechanism.

We close this Section by pointing out an unexpected difficulty arising
at the quantum-classical crossover. When reaching the crossover by decreasing 
the cutoff the advanced coordinates become suppressed. The semiclassical limit
corresponds to the dominance of the path integral by the configurations $\phi^\mr{adv}=0$.
But there is still another variable, $\phi$, to integrate over. Notice that
there is a cancellation between the actions $S[\phi^+]$ and $S^*[\phi^-]$ in the 
exponent on the right hand side of Eq. \eq{piwdefe} when $\phi^\mr{adv}=0$ causing the 
integrand of the path integral to become flat. This appears in the renormalization group scheme 
as a strong coupling regime for the configurations $\phi$. Thus, decoherence
induces strong coupling behavior and our approximations obtained within the
framework of the loop-expansion break down. Even the degrees of freedom may change.

This scenario is in strong contrast to the semiclassical limit discussed usually
in the framework of the path integral formalism. The limit $\hbar\to0$ of the
single time axis non-relativistic path integral
\be\label{trpi}
\int D[\phi]e^{\frac{i}{\hbar}S[\phi]}
\ee
is treated by the stationary phase approximation arguing that the
integral is dominated by the paths close to the classical trajectory. This description
applies to a closed quantum system which is not realistic in the macroscopic 
limit. When a density matrix is introduced to describe the separation of the system
from its environment, the time axis is doubled and the CTP path integral
\be\label{trpictp}
\int D[\phi^+]D[\phi^-]e^{\frac{i}{\hbar}S[\phi^+]-\frac{i}{\hbar}S^*[\phi^-]
+\frac{i}{\hbar}S_\mr{infl}[\phi^+,\phi^-]}
\ee
is used where $S_\mr{infl}[\phi^+,\phi^-]$ denotes the influence functional of the environment.
This expression is not dominated anymore by the classical trajectory. Instead there 
is an approximate cancellation between the two time axis when decoherence sets in, 
$S_\mr{infl}\approx M^2(\phi^+-\phi^-)^2/2$ with large $M^2$ and the physical field, 
$\phi$, is integrated over with approximately flat integrand.  The classical
limit is supposed to be recovered on the level of the equations of motion
for the expectation values of observables only. The approximation to the
transition amplitude \eq{trpi} in the semiclassical limit is based on the
rigidity of the dominant trajectory, the excitations being suppressed by $1/\hbar$.
In contrast of this simple picture the semiclassical limit is supposed to be
reached when the energy level density becomes high. This view takes into account
the environment and is supported by the strong coupling, soft saddle point
scenario of the functional integral \eq{trpictp} of the CTP formalism.

\section{Summary}\label{concl}
The dynamics of the expectation value of local operators in QED, the electric current 
and the electromagnetic field in particular, was studied in this work.
The expectation values were given in the framework of the CTP formalism covering
either fixed initial and final states or open, unconstrained time evolution.

It was shown that the equations of motion
for the expectation values of local operators can be obtained as variational
equations of a suitably defined effective action mixing the diagonal and off-diagonal 
quantum fluctuations in field diagonal basis. The effective action plays a double
role. It is not only a classical action for a classical field theory for the 
expectation values of local operators but it appears also as a Wilsonian 
effective action for degrees of freedom coupled to the system by our observables.

The effective actions for free photons and electrons are calculated first. While the 
former is a trivial exercise the latter proves to be a highly involved problem
due to the multi-particle aspects of the Dirac-see. The dynamics of
localized charged states depends on the states which contribute to the
current expectation value in a manner reminiscent of small and large polarons of Solid State Physics.
On the one hand, negative energy single-particle states contributing to the 
polarization generated by weak external field overlap strongly. As a result of the
dynamics of two-particle states generated by the current from the vacuum,
Pauli-blocking generates a long-range correlation which
produces separation-independent forces preventing the separation of localized charges 
by weak external electromagnetic field in the non-interacting Dirac-see. 
On the other hand, the charges arising from localized states within the mass gap
decouple from the filled up negative energy states and from each other. For such 
charges the standard relativistic action for localized particles was recovered. 
The actual value of the mass parameter depends on the details 
of the bound state and it gives rise a renormalization condition on the classical field
theory level.

The effective action for the interacting electron-photon system was constructed
up to quadratic parts in the fields on the two-loop order. The elimination of
either the current or the electromagnetic field produces the one-loop
action at a distance theory. The quadratic part of the one-loop effective action
for the current and the electromagnetic field calculated in the leading order of 
the gradient expansion agrees with the action of classical electrodynamics.
Quantum corrections to the action come from three directions. 
Vacuum polarizations due to electron-photon
interactions generate form-factors and new, higher order contributions in 
the fields which can be treated perturbatively having $e^2$ as small parameter. 
Other $e$-independent quantum corrections appear due to the 
indistinguishability of electrons in the non-interacting Dirac-see. 
The small parameter to organize an expansion in the exchange
effects is the absolute magnitude of the ratio of the electron Compton wavelength 
square and the invariant length square of the separation in space-time.
Finally, the third class of contributions corresponds to the 
boundary conditions in time which are build in at a microscopic level. 

The Dirac-see is the natural polarizable medium of QED and its effects can be taken 
into account in the same manner as in classical electrodynamics, by the 
introduction of the deplacement fields and dispersive electric permeability, 
when the action of the classical field theory is truncated in the quadratic 
approximation. 

Though the effective actions mentioned in this work govern expectation values
of a single measurement only they give some qualitative informations as 
far as multiple non-demolishing measurements are concerned. The correlation 
function of two successive non-demolishing measurements of the current is
the inverse of the $O(J^2)$ kernel of the effective action. In general, 
the multiple current measurement results are weakly correlated by 
vacuum-polarization effects when they are carried out in space-time
locations separated by an invariant lengths whose absolut magnitude exceeds the 
Compton wavelength.

The CTP formalism is particularly well suited to trace the way the quantum boundary 
conditions in time generate the time arrow in classical physics. A conjecture was put 
forward, namely that
closed quantum systems maintain time reversal invariance for any boundary conditions 
in time. When the system becomes open by coupling it to external sources
the response is given in terms of retarded Green functions and a time arrow is
generated for unconstrained, open ended time evolution. 
It was found that advanced effects mix in at some order only if the system is not 
allowed to follow an open ended time evolution and is projected on 
some states at the final time. Advanced effects represent
an influence of the future on the present and are possible in Quantum Mechanics
due to the superposition principle which allows us to specify both initial and final 
states. The symmetrical treatment of two time axis with opposite sense of time 
in the CTP highlights the fact that such effects are as natural as the influence of the
initial condition on the course of the motion. Similar phenomenon exists in classical physics, as well.
By specifying the initial and final coordinates of a classical system the change 
of the final coordinate modifies the whole time evolution. But this is a triviality
because the state of the system became uniquely defined at the final point only,
the equations of motion being of second order. Notice that the influence of the
boundary conditions in time does not decrease when the distance in time between
an observation and the boundary conditions is increased, in other words
there is no clusterisation in time due the unitarity of the time evolution.

The CTP formalism duplicates each degree of freedom, $\phi\to\phi^\pm$, 
for the treatment of the density matrix $\rho(\phi^+,\phi^-)$ and the two copies are 
subject to opposite time arrow. The sum $\phi^++\phi^-$ 
(difference $\phi^+-\phi^-$) remains invariant (changes sign) under the inversion of the time and follows
retarded (advanced) dynamics. The observed averages are given by 
a certain linear combination of $\phi^+$ and $\phi^-$ and classical field 
theories constructed for the measured fields only can not identify the sense of 
time. As a result, the dynamics of such theories is time reversal invariant by
mixing retarded and advanced effects with equal weight. 
The retarded and advanced effects can be separated by retaining both $\phi^+$ and 
$\phi^-$ in the classical dynamics. The resulting dynamics reflects decoherence 
on the average because $\la\Psi|\phi^+-\phi^-|\Psi\ra=0$. 
The classical field theory constructed for both $\phi^+$ and $\phi^-$ predicts
indeed retarded dynamics for the measured combination.
Decoherence, the suppression of $\phi^+-\phi^-$, generates macroscopical time arrow
by suppressing degrees of freedom subject to advanced dynamics. 

The effective action formalism offers classical dynamics both for the
microscopic and macroscopic regimes because the space-time resolution of 
the expectation values is limited by the UV cutoff $\Lambda$.
For high enough $\Lambda$ we see the microscopic quantum structure of local operators 
and the effective action is an extension of the density functional theory. 
As the cutoff is lowered one recovers a hydrodynamics description, classical
field theory for the current, and macroscopic physics. The issue of quantum-classical 
crossover can be addressed in this generalization of the renormalization group scheme
for the density matrix. The decoherence is
specially clear in the CTP formalism where it appears as a simple one-loop effect.
It is conjectured that the quantum-classical crossover which is characterized by 
decoherence is actually a strong coupling phenomenon for the physical fields. 

The qualitative similarity between suppression of advanced
combination of the degrees of freedom and decoherence suggests that the 
equations of motions of classical field theories which trace both degrees
of freedom of the CTP formalism changes less at the quantum-classical crossover
than those of the classical theories containing the measurable fields only.
Thus the equations of motion for the measurable field interpolate between the
time symmetrical form of the classical theory for the physical fields only
and the retarded equations of the classical theory for all fields as $\Lambda$
is decreased and cuts through the quantum-classical crossover.
This reasoning shows the peculiarity of the Abraham-Lorentz force of radiation damping.
This force originates from the consistent motion of all charges, interacting by 
the near-field Li\`enard-Wiechert potentials when the space-time resolution
of the expectation values is microscopic. But once the space-time resolution is 
chosen to be macroscopic the Abraham-Lorentz force arises from the accelerating 
"blocked" charge alone. Such a view of radiation damping applies to dissipative
forces in general and explains the need of auxiliary variables when such effects
are incorporated into canonical dynamics. Another remarkable fact is that the 
quadratic part of the action for the physical fields is independent of the 
boundary conditions in time. Consider two Universes corresponding to 
the same initial conditions but one is allowed to have an open, unconstrained time 
evolution while the state of the other is projected on the vacuum at a final time. 
The latter represents a completely absorbing Universe and its linearized equations 
of motions are identical with that of the Universe with open ended time evolutiuon.

The application of Quantum Field Theory to establish the dynamics of the
expectation values offers a comprehensive framework for measurement theory.
The basic difficulties of the measurement theory is to identify well established
and understood quantum effects in the interaction of a small and a large system. The
difficulty comes from the presence of a dimensionless number, $n/N$, the ratio of 
the number of degrees of freedom in the two systems taking extremely small values
and producing unexpected effects in a manner analogous to Statistical Physics. 
The $n/N$-dependence of the expectation values
can be traced by the study of the cutoff-dependence because the small system can
be imagined as residing in the elementary volume element of the underlying
field theory model. The cutoff-dependence can systematically be obtained by
eliminating the degrees of freedom, by the application of the renormalization group 
strategy based on a suitable chosen blocking procedure. 
The quantum-classical crossover shows some formal similarity with
spontaneous symmetry breaking, both representing new, less symmetrical scaling 
laws in the infrared scaling regime. The radiational time arrow is passed between
subsystems by this mechanism. In the case of QED the time reversal invariance
for the charges is broken by the boundary conditions in time imposed for photons
at the quantum-classical crossover. This mechanism is realized by the imaginary part of the photon 
propagator which is non-vanishing for continuous spectrum, ie. the infinite time limit
only. Another, conjectured feature of the quantum-classical crossover, the strongly coupled 
dynamics, might be part of the problem which prevents 
us from having a better insight into the quantum-classical transition.

\acknowledgments
This work could not have been carried out without the stimulating discussions
with Janos Hajdu whose critical reading of the manuscript was helpful, too. 
Furthermore I thank Kornel Sailer for the careful check of some formulae and
Andor Frenkel for pointing out the relevance of refs. \cite{ahare,ahark}.

\appendix
\section{Two-point function for hermitian scalar field}\label{freegenf}
In this Appendix the generating functional is constructed for a hermitian scalar 
subject to boundary conditions $\rho_i=|0\ra\la0|$ and $\rho_f=\openone$.
We start by collecting some formulae for the propagators of any hermitian, 
spinless, local operator $\phi$. The retarded, advanced near and far field
Green functions are introduced in the next step, followed by a
brief recall of the spectral representation. Finally, the propagator of a free,
neutral scalar particle is discussed and the massless results are given 
in closed form.

\subsection{Propagator as a block-matrix}
The connected Green functions are obtained by means of the generating functional 
\be
e^{iW[\hj]}=\Tr\at[e^{i\int_{t_i}^{t_f}dt'j^-(t')\phi_-(t')]}]
T[e^{i\int_{t_i}^{t_f}dt'j^+(t')\phi_+(t')}]|0\ra\la0|
\ee
and the propagator is defined as a $2\times2$ block-matrix
\be
\Delta^{\sigma\sigma'}_{xy}=-\fdd{W[\hj]}{j^\sigma_x}{j^{\sigma'}_y}_{|\hj=0}.
\ee
When both legs of the propagator are on the positive-sense time axis one finds the 
usual causal propagator
\bea
i\Delta^{++}_{xy}&=&\sum_n\la n|T[\phi_x\phi_y]|0\ra\la0||n\ra\nn
&=&\la0|T[\phi_x\phi_y|0\ra.
\eea
The propagator along the negative-sense time axis is
\bea
i\Delta^{--}_{xy}&=&\sum_n\la n||0\ra\la0|\at[\phi_x\phi_y]|n\ra\nn
&=&\la0|T[\phi_y\phi_x]|0\ra^*\nn
&=&(i\Delta^{++}_{yx})^*.
\eea
Finally, the mixed propagator is given by
\be
i\Delta^{+-}_{xy}=\sum_n\la0|\phi_y|n\ra\la n|\phi_x|0\ra.
\ee
The choice $\rho_f=\openone$ allows us to decrease $t_f$ until it reaches
$\max(x^0,y^0)$ and placing both field operators into the same expectation
value resulting 
\be
i\Delta^{+-}_{xy}=\la 0|\phi_y\phi_x|0\ra.
\ee

Note that the identity $T[\phi_x\phi_y]+\at[\phi_x\phi_y]=\phi_x\phi_y+\phi_y\phi_x$ 
leads to the relation
\be\label{dpmsum}
\Delta^{++}+\Delta^{--}=\Delta^{+-}+\Delta^{-+}.
\ee

\subsection{Retarded and advanced propagators}
Introducing the notation $\Delta=\Delta^{++}$ for the causal propagator
\be
i\Delta_{x,x'}=\Theta(t-t')\la0|\phi_x\phi_{x'}|0\ra+\Theta(t'-t)\la0|\phi_{x'}\phi_x|0\ra,
\ee
its real an imaginary parts are
\be
\Re\Delta_{x,x'}=-\hf\Theta(t-t')i\la0|[\phi_x,\phi_{x'}]|0\ra
-\hf\Theta(t'-t)i\la0|[\phi_{x'},\phi_x]|0\ra
=-\hf\epsilon(t-t')i\la0|[\phi_x,\phi_{x'}]|0\ra
\ee
with $\epsilon(t)=\mr{sign}(t)$ and
\be
\Im\Delta_{x,x'}=-\hf\Theta(t-t')\la0|\{\phi_x,\phi_{x'}\}|0\ra
-\hf\Theta(t'-t)\la0|\{\phi_{x'},\phi_x\}|0\ra
=-\hf\la0|\{\phi_x,\phi_{x'}\}|0\ra,
\ee
respectively. The retarded and advanced propagators are defined as usual,
\be
i\Delta^{\stackrel{r}{a}}_{xx'}=\pm\Theta(\pm(t-t'))\la0|[\phi_x,\phi_{x'}]|0\ra.
\ee
It is easy to find the actual expressions in terms of $\hDD$ by noting
\bea
i\Delta^r&=&\Theta(t-t')\la0|\phi_x\phi_{x'}|0\ra+\Theta(t-t')\la0|\phi_{x'}\phi_x|0\ra\nn
&=&i\Delta_{xx'}-\Theta(t'-t)\la0|\phi_{x'}\phi_x|0\ra-\Theta(t-t')\la0|\phi_{x'}\phi_x|0\ra\nn
&=&i\Delta_{xx'}-i\Delta^{+-}_{xx'}\nn
&=&i\Delta^*_{xx'}+i\Delta^{-+}_{xx'},
\eea
where the relation $i\Delta^{--}=(i\Delta^{++})^*$ and Eq. \eq{dpmsum} were used in 
obtaining the last equation. The near and far field propagators,
$\Delta^n=\Re\Delta$ and $\Delta^f=2\Re\Delta^{-+}$ satisfy the equation
\be
\Delta^r=\Delta^n+\hf\Delta^f.
\ee
Similar reasoning yields the equation
\be
\Delta^a=\Delta^n-\hf\Delta^f
\ee
for the advanced propagator. The complete propagator can be parameterized by means of
three real functions, for instance
\be
\hDD=\begin{pmatrix}\Delta^n+i\Im\Delta&-\hf\Delta^f+i\Im\Delta\cr
\hf\Delta^f+i\Im\Delta&-\Delta^n+i\Im\Delta\end{pmatrix}
\ee

\subsection{Spectral representation}
A particularly useful parameterization of the propagator is provided by the spectral 
representation. By following the standard procedure introduce the eigenvectors of 
the energy-momenta, $P^a|n\ra=p^a_n|n$, to induce the space-time dependence
$\phi_x=e^{iPx}\phi_0e^{-iPx}$ and write 
\bea\label{terszorz}
i\la0|\phi_x\phi_y|0\ra&=&\sum_ni\la0|\phi_x|n\ra\la n|\phi_{x'}|0\ra\nn
&=&\sum_ni|\la0|\phi_0|n\ra|^2e^{-ip_n(x-x')}\nn
&=&i\int_pe^{-ip(x-y)}\Theta(p^0)\rho(p^2),
\eea
where
\be
\Theta(p^0)\rho(p^2)=\frac{1}{2\pi}\sum_n(2\pi)^4\delta(p_n-p)|\la0|\phi_0|n\ra|^2,
\ee
is the spectral function. One can recast the last line of Eqs. \eq{terszorz} in the form 
reminiscent of free propagators,
\bea
i\la0|\phi_x\phi_y|0\ra&=&i\int_p\int_0^\infty d\mu^2e^{-ip(x-x')}\Theta(p^0)\rho(\mu^2)2\pi\delta(p^2-\mu^2)\nn
&=&i\int_{\tilde{\v{k}}}\int_0^\infty d\mu^2
e^{-i\omega_\v{k}(\mu^2)(t-t')+i\v{k}(\v{x}-\v{x}')}\rho(\mu^2),
\eea
where $\omega_\v{k}(\mu^2)=\sqrt{\mu^2+k^2}$ and
\be
\int_{\tilde{\v{k}}}=\int\frac{d\v{k}}{(2\pi)^32\omega_{\mu^2,\v{k}}}.
\ee
The propagator $\Delta$ is obtained by means of the Fourier representation of the 
Heavyside-function,
\be\label{causale}
\Delta_{xx'}=-\int_0^\infty d\mu^2\rho(\mu^2)\int_{\tilde{\v{k}},\omega}\left(
\frac{e^{-i(\omega_\v{k}(\mu^2)+\omega)(t-t')+i\v{k}(\v{x}-\v{x'})}}{\omega+i\epsilon}
+\frac{e^{i(\omega_\v{k}(\mu^2)+\omega)(t-t')-i\v{k}(\v{x}-\v{x'})}}{\omega+i\epsilon}\right).
\ee
and performing the change of variable $\omega\to-\omega$ in the second integral
\bea\label{causalk}
\Delta_{xx'}&=&-\int_0^\infty d\mu^2\rho(\mu^2)\int_{\tilde{\v{k}},\omega}\left(
\frac{e^{-i(\omega_\v{k}(\mu^2)+\omega)(t-t')+i\v{k}(\v{x}-\v{x'})}}{\omega+i\epsilon}
+\frac{e^{i(\omega_\v{k}(\mu^2)-\omega)(t-t')+i\v{k}(\v{x}-\v{x'})}}{-\omega+i\epsilon}\right)\nn
&=&\int_0^\infty d\mu^2\rho(\mu^2)\Delta_{0~xx'}(\mu^2)
\eea
where
\be\label{freecapr}
\Delta_{0~xx'}(\mu^2)=-\int_k\frac{e^{-ik(x-x')}}{k^2-\mu^2+i\epsilon}.
\ee
The steps above repeated for the block-propagator yield
\be
\hDD=\int_0^\infty d\mu^2\rho(\mu^2)\hDD_0(\mu^2).
\ee

\subsection{Free neutral scalar particle}
The path integral representation of non-interacting generating functional is
\bea
e^{iW_0[\hj]}&=&\Tr\at[e^{i\int_{t_i}^{t_f}dt'[H(t')+\int_\v{x}j^-_{t',\v{x}}\phi^-_{t',\v{x}}]}]]
T[e^{-i\int_{t_i}^{t_f}dt'[H(t')-\int_\v{x}j^+_{t',\v{x}}\phi^+_{t',\v{x}}]}]|0\ra\la0|\nn
&=&\int D[\hat\phi]e^{\frac{i}{2}\hat\phi\cdot\hDD^{-1}_0\cdot\hat\phi+i\hj\cdot\hat\phi}.
\eea
where $\hDD_0=\hDD_0(m^2)$. The path integral can be carried out with the result
\be
W_0[\hj]=-\hf\hj\cdot\hDD_0\cdot\hj.
\ee

The detailed form of the propagators can be obtained by means of the Fourier integral
\be
\phi_x=\int_{\tilde{\v{k}}}
[e^{-i\omega_\v{k}t+i\v{k}\v{x}}a_\v{k}+e^{i\omega_\v{k}t-i\v{k}\v{x}}a^\dagger_\v{k}]
\ee
where $\omega_\v{k}=\omega_\v{k}(m^2)$. The non-vanishing canonical commutation 
relation are $[a_\v{k},a^\dagger_\v{k'}]=(2\pi)^32\omega_\v{k}\delta_{\v{k},\v{k'}}$. 
The repetition of the steps \eq{causale}-\eq{causalk} leads to the causal propagator \eq{freecapr}.

\subsection{Real space expressions for $m=0$}
First we calculate the expectation value
\bea
\la0|\phi_x\phi_{x'}|0\ra&=&\int_{\tilde{\v{k}}}e^{-ik(x-x')}\nn
&=&\frac{1}{8\pi^2}\int_0^\infty dkk\int_{-1}^1 d\cos\theta e^{-ikt+ikr\cos\theta}.
\eea
The contributions to the $k$-integration are negligible at the upper limit when working 
in a cutoff theory which can be incorporated by calculating the integral for
$t\to t-i\epsilon$ yielding
\be
\la0|\phi_x\phi_{x'}|0\ra=-\frac{1}{4\pi^2}\frac{1}{(x-x')^2-i\epsilon(x^à-x^{'0})\epsilon}
\ee
and
\be\label{dorsp}
D_{0~xx'}=\frac{1}{4\pi}\delta((x-x')^2)-\frac{i}{4\pi^2}P\frac{1}{(x-x')^2},
\ee
$D_0=\Delta_0(0)$ being the free massless propagator. 
The retarded and advanced propagators are
\be
D^{\stackrel{r}{a}}_{0~xx'}=\frac{1}{2\pi}\Theta(\pm(x^0-x^{0'}))\delta((x-x')^2).
\ee

\section{Two-point vertex function for non-interacting Dirac-see}\label{qefnds}
In this Appendix we present the calculation of the kernel of the 
quadratic effective action for the current with $\beta=0$ for the non-interacting 
Dirac-see. 

The Fourier representation of the quadratic form \eq{qfreactk}
\be
\Gamma^{(2)\mr{el}}_{(t_1,\v{x}_1),(t_2,\v{x}_2)}
=-\frac{15\pi m^2}{(2\pi)^4}[\Theta(t_1-t_2)+\Theta(t_2-t_1)]
\int_\v{k}e^{-i\v{k}(\v{x}_1-\v{x}_2)}\int_{k_0}\frac{e^{ik_0(t_1-t_2)}}{(k_0^2-\v{k}^2+i\epsilon)^2}
\ee
gives after straightforward integration over the frequency and the polar angles
\be\label{kpvf}
\epsilon(t)\Gamma^{(2)\mr{el}}(t,r)=\frac{15m^2}{8r}[itf_1(t+r)-itf_1(t-r)+f_2(t+r)-f_2(t-r)]
\ee
with $t=t_1-t_2$, $r=|\v{x}_1-\v{x}_2|$ and
\be
f_n(t)=\frac{1}{2\pi}\int_k\frac{e^{-ikt}}{k^n}.
\ee
The relation $\partial_t^nf_n(t)=(-i)^n\delta(t)$ yields
$f_1(t)=c_1-i\Theta(t)$ and $f_2(t)=-t\Theta(t)+tc_2+c_3$, $c_k$ being integration constants
to be determined. The vertex function
\be
\Gamma^{(2)\mr{el}}(t,r)=\frac{\epsilon(t)15m^2}{8}[2c_2-\Theta(t-r)-\Theta(t+r)]
\ee
obtained by replacing the solution for $f_1(t)$ and $f_2(t)$ into Eq. \eq{kpvf} 
should be symmetric in $t$ therefore $c=1/2$ and Eq. \eq{vfctrt} follows.

\section{Loop expansion for $W[\ha,\hj]$}\label{loopexp}
In this Appendix some details of the calculation of the generating functional for 
connected Green functions of the current and the photon field in QED are presented.

\subsection{Setting up the loop expansion}
We start by Eq. \eq{wqedfn} for the generating functional written after the photon field
is integrated out,
\be
e^{iW[\ha,\hj]}=e^{iW^\mr{el}[\ha+ie\hs\fd{}{\hj}]+iS_{CT}[-ie\fd{}{\hj}]}e^{-\frac{i}{2}\hj\cdot\hD_0\cdot\hj}.
\ee
The identity 
\be
\fd{}{f}e^{F[f]}\openone=e^{F[f]}\left(\fd{}{f}+\fd{F[f]}{f}\right)\openone
\ee
yields
\bea\label{wjale}
e^{iW[\ha,\hj]}&=&e^{-\frac{i}{2}\hj\cdot\hD_0\cdot\hj}
e^{iW^\mr{el}[\ha+ie\hs\fd{}{\hj}+e\hs\hD_0\cdot\hj]+iS_{CT}[-ie\fd{}{\hj}-e\hD_0\cdot\hj]]}\openone\nn
&=&e^{-\frac{i}{2}\hj\cdot\hD_0\cdot\hj}
\sum\limits_{m=0}^\infty\frac{i^m}{m!}\left[\sum_{n=0}^\infty\frac{i^n}{n!}
\left(e\fd{}{\hj}\hs\cdot\fd{'}{\ha'}\right)^nW^\mr{el}[\ha+\ha']\right]^m\openone
\eea
where 
\be\label{gencurr}
\ha'=e\hs\hD_0\cdot\hj
\ee
and $\delta'/\delta\ha'$ acts only on the first factor of $W^\mr{el}$. For simplicity the 
counterterms are suppressed. The $\hj$-dependence comes from $\ha'$ therefore we have
\be\label{pertexp}
e^{iW[\ha,\hj]}=e^{-\frac{i}{2}\hj\cdot\hD_0\cdot\hj}\sum\limits_{m=0}^\infty\frac{i^m}{m!}
\left[\sum_n\frac{1}{n!}\left(\fd{}{\ha'}\cdot ie^2\hs\hD_0\hs
\cdot\fd{'}{\ha'}\right)^nW^{e'}[\ha+\ha']\right]^m\openone.
\ee

\subsection{Two-loop order quadratic generating functional}
We shall calculate $W[\ha,\hj]$ on the two-loop order up to quadratic terms in the
external sources. To this end we need
\be\label{wpser}
W=W^\mr{el}[\ha+\ha']+\fd{}{\ha'}\cdot ie^2\hs\hD_0\hs\cdot\fd{'}{\ha'}W^\mr{el}[\ha+\ha']
+\hf\left(\fd{}{\ha'}\cdot ie^2\hs\hD_0\hs\cdot\fd{'}{\ha'}\right)^2W^\mr{el}[\ha+\ha']
\ee
and its square in Eq. \eq{pertexp} up to two-loop terms. We start with the expansion in $\ha'$,
\be
W^\mr{el}[\ha+\ha']\approx W^\mr{el}[\ha]+W^{\mr{el}(1)}_a[\ha]\ha'_a+\hf W^{\mr{el}(2)}_{ab}[\ha]\ha'_a\ha'_b
+\frac{1}{3!}W^{\mr{el}(3)}_{abc}[\ha]\ha'_a\ha'_b\ha'_c+\frac{1}{4!}W^{\mr{el}(4)}_{abcd}[\ha]\ha'_a\ha'_b\ha'_c\ha'_d
\ee
by retaining the $O(\ha^4)$ pieces only. Two functional derivations give
\bea
\fd{}{\ha'_a}W^\mr{el}[\ha+\ha']&\approx&W^{\mr{el}(1)}_a[\ha]+W^{\mr{el}(2)}_{ab}[\ha]\ha'_b
+\frac{1}{2}W^{\mr{el}(3)}_{abc}[\ha]\ha'_b\ha'_c+\frac{1}{3!}W^{\mr{el}(4)}_{abcd}[\ha]\ha'_b\ha'_c\ha'_d\nn
\fdd{}{\ha'_a}{\ha'_b}W^\mr{el}[\ha+\ha']&\approx&W^{\mr{el}(2)}_{ab}[\ha]
+W^{\mr{el}(3)}_{abc}[\ha]\ha'_c+\hf W^{\mr{el}(4)}_{abcd}[\ha]\ha'_c\ha'_d,
\eea
where the coefficients are
\bea
W^{\mr{el}(1)}[\ha]&\approx&\hj[\ha]=W^{\mr{el}(2)}\ha,\nn
W^{\mr{el}(2)}_{ab}[\ha]&\approx&W^{\mr{el}(2)}_{ab}+\hf W^{\mr{el}(4)}_{abcd}\ha_c\ha_d,\nn
W^{\mr{el}(3)}_{abc}[\ha]&\approx&W^{\mr{el}(4)}_{abcd}\ha_d,\nn
W^{\mr{el}(4)}_{abcd}[\ha]&\approx&W^{\mr{el}(4)}_{abcd}+\hf W^{\mr{el}(6)}_{abcdef}\ha_e\ha_f
\eea
in the given order. The photon exchange appears in the combinations
\bea
W^{\mr{el}(2)}_D[\ha]&=&W^{\mr{el}(2)}_{ab}[\ha](\hs i\hD_0\hs)_{ab}\nn
W^{\mr{el}(3)}_D[\ha]_a&=&W^{\mr{el}(3)}_{abc}[\ha](\hs i\hD_0\hs)_{bc}\nn
W^{\mr{el}(4)}_D[\ha]_{ab}&=&W^{\mr{el}(4)}_{abcd}[\ha](\hs i\hD_0\hs)_{cd}.
\eea
The expressions
\bea
\fd{}{\ha'}\cdot i\hs\hD_0\hs\cdot\fd{'}{\ha'}W^\mr{el}[\ha+\ha']
&\approx&W^{\mr{el}(2)}_D[\ha]+W^{\mr{el}(3)}_D\cdot\ha'+\hf\ha'\cdot W^{\mr{el}(4)}_D[\ha]\cdot\ha'\nn
&&\hskip-40pt+\left(W^{\mr{el}(1)}_a[\ha]+W^{\mr{el}(2)}_{ab}[\ha]\ha'_b+\frac{1}{2}W^{\mr{el}(3)}_{abc}[\ha]\ha'_b\ha'_c
+\frac{1}{e3!}W^{(4)}_{abcd}[\ha]\ha'_b\ha'_c\ha'_d\right)(\hs i\hD_0\hs)_{ae}\fd{}{\ha'_e},\nn
\left(\fd{}{\ha}\cdot\hs i2\hD_0\hs\cdot\fd{'}{\ha'}\right)^2W^\mr{el}[\ha+\ha']
&\approx&\left(W^{\mr{el}(3)}_{Da}[\ha]+W^{\mr{el}(4)}_{Dab}[\ha]\ha'_b\right)(\hs i\hD_0\hs)_{ae}\fd{}{\ha'_e}.
\eea
give
\bea
W&=&W^\mr{el}[\ha]+W^{\mr{el}(1)}[\ha]\cdot\ha'+\hf\ha'\cdot W^{\mr{el}(2)}_{ab}[\ha]\cdot\ha'
+\frac{1}{3!}W^{\mr{el}(3)}_{abc}[\ha]\ha'_a\ha'_b\ha'_c
+\frac{1}{4!}W^{\mr{el}(4)}_{abcd}[\ha]\ha'_a\ha'_b\ha'_c\ha'_d\nn
&&+e^2W^{\mr{el}(2)}_D[\ha]+e^2W^{\mr{el}(3)}_D\cdot\ha'+\frac{e^2}{2}\ha'\cdot W^{\mr{el}(4)}_D[\ha]\cdot\ha'
\eea
in the given truncation. Its square
\be
W^2\approx\left(W^\mr{el}[\ha]+W^{(1)}_a[\ha]\ha'_a+\hf W^{(2)}_{ab}[\ha]\ha'_a\ha'_b\right)^2
+e^2(\ha+\ha')\cdot W^{(2)}[\ha]\cdot\hs i\hD_0\hs\cdot W^{(2)}[\ha]\cdot(\ha+\ha')
\ee
is obtained by taking into account unitarity, $W^\mr{el}[0]=0$. We have now everything 
needed to re-exponentiate the necessary terms in Eq. \eq{pertexp},
\bea
1+iW-\hf W^2&\approx&\exp i\biggl[W^\mr{el}[\ha]+W^{\mr{el}(1)}[\ha]\cdot\ha'+\hf\ha\cdot W^{\mr{el}(2)}[\ha]\cdot\ha'
+e^2W^{\mr{el}(2)}_D[\ha]+e^2W^{\mr{el}(3)}_D\cdot\ha'+\frac{e^2}{2}\ha'\cdot W^{\mr{el}(4)}_D[\ha]\cdot\ha'\nn
&&-\frac{e^2}{2}(\ha+\ha')\cdot W^{\mr{el}(2)}[\ha]\cdot\hs\hD_0\hs\cdot W^{\mr{el}(2)}[\ha]\cdot(\ha+\ha')]\biggr]
\eea
yielding
\bea
W[\ha,\hj]&=&-\hf\hj\cdot\hD_0\cdot\hj-\hf(\ha+\ha')\cdot\htG_0\cdot(\ha+\ha')
+\frac{e^2}{2}W^{\mr{el}(4)}_{abcd}\ha_c\ha_d(\hs i\hD_0\hs)_{ab}\nn
&&+e^2W^{\mr{el}(4)}_{abcd}\ha'_a\ha_d(\hs i\hD_0\hs)_{bc}
+\frac{e^2}{2}W^{\mr{el}(4)}_{abcd}(\hs i\hD_0\hs)_{cd}\ha'_a\ha'_b
-\frac{e^2}{2}(\ha+\ha')_a\htG_{0ab}(\hs\hD_0\hs)_{be}\htG_{0ef}(\ha+\ha')_f.
\eea
The substitution of $\ha'$ given by Eq. \eq{gencurr} results in
\bea
W[\ha,\hj]&=&\hf\hj\cdot[-\hD_0-e^2\hD_0\hs\cdot(\htG_0-e^2W^{(4)}_D)\cdot\hs\hD_0
-e^4\hD_0\hs\cdot\htG_0\cdot\hs\hD_0\hs\cdot\htG_0\cdot\hs\hD_0]\cdot\hj\\
&&+\hf\ha(-\htG_0+e^2W^{(4)}_D-e^2\htG_0\cdot\hs\hD_0\hs\cdot\htG_0)\ha
+e\hj[-\hD_0\hs\cdot(\htG_0-e^2W^{(4)}_D)-e^2\hD_0\hs\cdot\htG_0\cdot\hs\hD_0\hs\cdot\htG_0]\ha\nonumber
\eea
leading to expressions \eq{ctpw}-\eq{ctppse} for the generating functional. The 
parameterization \eq{jpar} and \eq{apar} give
\be\label{wrepk}
\Re W[a,\ab,j,\jb]=\hf(a,j,\ab,\jb)\cdot
\begin{pmatrix}\frac{\kappa}{2}(K+K^\mr{tr})&K\cr K^\mr{tr}&0\end{pmatrix}
\cdot\begin{pmatrix}a\cr j\cr\ab\cr\jb\end{pmatrix}
\ee
with
\be
K_\mr{OBC}=-\begin{pmatrix}\tG^\mr{ret}&e\tG^\mr{ret}D_0^\mr{ret}\cr eD_0^\mr{ret}\tG^\mr{ret}&D^\mr{ret}\end{pmatrix}
\ee
for OBC shown in a detailed form in Eq. \eq{inwobc}. For FBC the sources belonging 
to different time axis decouple giving Eq. \eq{wrepk} with
\be
K_\mr{FBC}=-\begin{pmatrix}\tG^\mr{near}&e(\tG^\mr{near}D_0^\mr{near}+\Im\tG\Im D_0)\cr
e(D_0^\mr{near}\tG^\mr{near}+\Im D_0\Im\tG)&D_0\end{pmatrix}.
\ee

For the sake of completeness we record the imaginary part,
\be\label{wrepko}
\Im W_\mr{OBC}[a,j]=-\hf(a,j)\cdot\begin{pmatrix}\Im\tG&e(\Im\tG D_0^\mr{adv}+\tG^\mr{ret}\Im D_0)\cr
e(D_0^\mr{ret}\Im\tG+\Im D_0\tG^a)&\Im D\end{pmatrix}\cdot\begin{pmatrix}a\cr j\cr\end{pmatrix},
\ee
and
\be\label{wrepkf}
\Im W_\mr{FBC}[a,\ab,j,\jb]=-\hf(a,\ab,j,\jb)\cdot\begin{pmatrix}\Im\tG\begin{pmatrix}\frac{1+\kappa^2}{2}&\kappa\cr\kappa&2\end{pmatrix}&
e(\Im\tG D_0^\mr{near}+\tG^\mr{near}\Im D_0)\begin{pmatrix}\frac{1+\kappa^2}{2}&\kappa\cr\kappa&2\end{pmatrix}\cr
e(D_0^\mr{near}\Im\tG+\Im D_0\tG^\mr{near})\begin{pmatrix}\frac{1+\kappa^2}{2}&\kappa\cr\kappa&2\end{pmatrix}&
\Im\hD\begin{pmatrix}\frac{1+\kappa^2}{2}&\kappa\cr\kappa&2\end{pmatrix}
\end{pmatrix}\cdot\begin{pmatrix}a\cr\ab\cr j\cr\jb\end{pmatrix}.
\ee

\section{Legendre transformation}\label{qedgam}
The calculation of the quadratic effective action is sketched in this Appendix. 
There is a useful relation between the second derivatives of a function and its
Legendre transform. Let us denote all sources and fields in the problem by 
$j$ and $\phi$. Then the effective action is defined by $\Gamma[\phi]=W[j]-j\cdot\phi$
with $\phi=\delta W[j]/\delta j$. One can easily prove the identity
\be\label{wgmder}
\fdd{W[j]}{j}{j}\cdot\fdd{\Gamma[\phi]}{\phi}{\phi}=-\openone
\ee
which renders the calculation of the Legendre transform of a quadratic 
expression to an inversion problem.

We start with a few useful relations for the inversion of block-matrices to be used 
later and rules for obtaining the retarded and advanced components
from a product of propagators. This is followed by the calculation of the 
effective action for four fields with open and fixed boundary conditions.

\subsection{Inversion of block-propagators}
We shall meet two kinds of inversion during the Legendre transformation
of different functionals. One of them is the inversion of a $2\times2$ block propagator
\be
K=-\begin{pmatrix}\tG&eS_1\cr eS_2&D\end{pmatrix}.
\ee
The inverse
\be\label{invbm}
\begin{pmatrix}-\tG&-eS_1\cr-eS_2&-D\end{pmatrix}^{-1}
=-\begin{pmatrix}\tG^{\mr{int}-1}&-e\tG^{-1}S_1D^{\mr{int}-1}\cr-eD^{\mr{int}-1}S_2\tG^{-1}&D^{\mr{int}-1}\end{pmatrix}
\ee
where 
\be\label{tgc}
\tG^\mr{int}=\tG-e^2S_1D^{-1}S_2
\ee
and
\be\label{dc}
D^\mr{int}=D-e^2S_2\tG^{-1}S_1
\ee
is obtained by solving the system of linear equations
\be
\begin{pmatrix}a\cr b\end{pmatrix}=K\begin{pmatrix}x\cr y\end{pmatrix}
\ee
by elimination. 

Anther inversion problem one encounters involves the block-matrix
\be 
\begin{pmatrix}\frac{\kappa}{2}(K+K^\mr{tr})&K\cr K^\mr{tr}&0\end{pmatrix}.
\ee
The same strategy as in the previous case ie. the solution of the corresponding set of 
linear equations gives
\be \label{kinv}
\begin{pmatrix}\frac{\kappa}{2}(K+K^\mr{tr})&K\cr K^\mr{tr}&0\end{pmatrix}^{-1}
=\begin{pmatrix}0&K^{\mr{tr}-1}\cr K^{-1}&-\frac{\kappa}{2}(K^{\mr{tr}-1}+K^{-1})\end{pmatrix}.
\ee

\subsection{Light-cone as a null-space}\label{raton}
The real part of massless particle propagators are non-vanishing on the light-cone only.
This property simplifies enormously the expressions of the perturbation expansion
frequently involving a chain of products of massless propagators
$\Pi=\hD_1\hs\cdot\hD_2\hs\cdots\hD_n$ and its retarded and advanced parts.

We start with the demonstration of the useful relation
\be\label{dfdnme}
D^\mr{far}\cdot D^{\mr{near}-1}=0
\ee
valid for massless propagators. The straightforward proof of Eq. \eq{dfdnme}
is by inserting the propagator
\be
D^{\stackrel{\mr{ret}}{\mr{adv}}}_{xx'}=-\int_k\frac{e^{-ik(x-x')}}{(k^0\pm i\epsilon)^2-\v{k}^2}F(k^2),
\ee
with $F(0)=1$ resulting
\be
(D^\mr{far}\cdot D^{\mr{near}-1})_{xx'}=-i\epsilon\int_kPe^{-ik(x-x')}\frac{\epsilon(k^0)}{k^2}
\ee
which is indeed vanishing for $\epsilon\to0$. 

To understand better the regularization effects of $\epsilon$ we note that on the 
one hand $D_0=(D^\mr{ret}+D^\mr{adv})/2$ satisfies the inhomogeneous equation of motion
\be\label{fge}
\frac{1}{F(-\Box)}\cdot\Box\cdot D_0=\openone.
\ee 
Thus, $G^\mr{near}\to\infty$ and $G^{\mr{near}-1}\to0$ on the light cone for $\epsilon\to0$.
On the other hand, $D^\mr{far}$ is non-vanishing on the light cone only.
Therefore, Eq. \eq{dfdnme} follows. A more explicit argument is based on the fact that
$D^\mr{far}$ is a homogeneous Green function satisfying the equation 
\be
\frac{1}{F(-\Box)}\cdot\Box\cdot D^\mr{far}=0
\ee 
which is just Eq. \eq{dfdnme} due to Eq. \eq{fge}.

The lesson is the identity $D^{\mr{near}-1}\cdot D^\mr{ret}=D^{\mr{near}-1}\cdot D^\mr{adv}$, ie. the equation $D^\mr{ret}=D^\mr{adv}$
holds when the propagators appear in a product beside of a factor $D^{\mr{near}-1}$. 
This result yields 
\be\label{nifegys}
D^{\mr{near}-1}\cdot D^\mr{ret}=D^{\mr{near}-1}\cdot D^\mr{adv}=\openone.
\ee

Let us finally consider the product $\Pi$ where each propagator is of the form
\be\label{genformbpr}
D_j=\begin{pmatrix}0&D_j^\mr{adv}\cr D_j^\mr{ret}&i\Im D_j\end{pmatrix}
\ee
in the basis $(j^+-j^-,j^++j^-)$. It is easy to verify that $\Pi$ is of this form, as well, and
\bea
\hD_1\hs\cdots\hD_n&=&\begin{pmatrix}0&D^\mr{adv}_1\cdot D^\mr{adv}_2\cdots 
D^\mr{adv}_n\cr D^\mr{ret}_1\cdot D^\mr{ret}_2\cdots D^\mr{ret}_n&0\end{pmatrix}\nn
&&+\begin{pmatrix}0&0\cr0&\Im D_1\cdot D^\mr{adv}_2\cdots D^\mr{adv}_n
+D^\mr{ret}_2\cdot\Im D_2\cdots D^\mr{adv}_n\cdots+D^\mr{ret}_1\cdot D^\mr{ret}_2\cdots\Im D\end{pmatrix}
\eea
giving
\bea\label{raszorz}
(\hD_1\hs\cdots\hD_n)^r&=&D^\mr{ret}_1\cdots D^\mr{ret}_n,\nn
(\hD_1\hs\cdots\hD_n)^a&=&D^\mr{adv}_1\cdots D^\mr{adv}_n.
\eea
The property \eq{nifegys} can be used to show
\be
\hf(D_1^\mr{ret}\cdots D_n^\mr{ret}+D_1^\mr{adv}\cdots D_n^\mr{adv})=D_1^\mr{near}\cdots D_n^\mr{near}
\ee
giving
\be\label{nszernn}
(\hD_1\hs\cdots\hD_n)^\mr{near}=D^\mr{near}_1\cdots D^\mr{near}_n.
\ee
The fact that the product $\hD_1\hs\cdots\hD_n$ preserves the form \eq{genformbpr} 
assures that $W^{\mr{el}(4)}_D$ which appears in the self energy is of this form, too.

Eqs. \eq{raszorz} and \eq{nszernn} can be used to show that once the pair creation is
excluded ($\Im\tG=0$) the generating functional $W_\mr{FBC}[\ha,\hj]$ is 
obtained from $W_\mr{OBC}[\ha,\hj]$ by the simple rule found
for non-interacting systems, namely the replacement of all retarded or advanced propagator 
by their near field version.

\subsection{Effective actions for four fields}\label{efff}
To obtain the effective action for the fields $J$, $J^\mr{adv}$, $A$ and $A^\mr{adv}$
we need the inverse of the block-matrices of Eqs. \eq{inwobc} and \eq{inwfbc}.
We start with OBC when the two-loop expression electron self energy
in Eq. \eq{ctpese} and the first equation in \eq{raszorz} give
\be
\tG^\mr{ret}=G_0^\mr{ret}+e^2\tG_0^\mr{ret}\cdot D_0^\mr{ret}\cdot\tG_0^\mr{ret}-e^2W^{\mr{el}(4)\mr{ret}}_D.
\ee
In a similar manner the photon self energy of Eq. \eq{ctppse} gives
\be
D^\mr{ret}=\hD_0^\mr{ret}+e^2D_0^\mr{ret}\cdot(\tG_0^\mr{ret}-e^2W^{\mr{el}(4)\mr{ret}}_D)\cdot D_0^\mr{ret}
+e^4D_0^\mr{ret}\cdot\tG_0^\mr{ret}\cdot D_0^\mr{ret}\cdot\tG_0^\mr{ret}\cdot D_0^\mr{ret}
\ee
the inverse being
\be
D^{\mr{ret}-1}=D^{\mr{ret}-1}_0-e^2(\tG_0^\mr{ret}-W^{\mr{el}(4)\mr{ret}}_D).
\ee
These expressions substituted in Eqs. \eq{tgc} and \eq{dc} give
$D^\mr{int}=\hD_0^\mr{ret}$ and $\tG^\mr{int}=G_0^\mr{ret}-e^2W^{\mr{el}(4)\mr{ret}}_D$. The inverse \eq{invbm} gives 
the expression \eq{kinvo} when choosing $S_1=\tG^\mr{ret}\cdot D_0^\mr{ret}=S_2^\mr{tr}$. 
Finally, the effective action \eq{efafvo} follows from Eqs. \eq{wgmder} and \eq{kinv}
and its equations of motion give
\bea\label{eqmoff}
\tG^a_0a&=&-(1+e^2W^{\mr{el}(4)a}_D\cdot\tG^{a-1}_0)\cdot J^\mr{adv}+e\tG^a_0\cdot A^\mr{adv},\nn
D^\mr{adv}_0j&=&eD^\mr{adv}_0\cdot J^\mr{adv}-A^\mr{adv},\nn
\tG^\mr{ret}_0\cdot\left(\ab+\frac{\kappa}{2}a\right)&=&-(1+e^2W^{\mr{el}(4)\mr{ret}}_D\cdot\tG^{\mr{ret}-1}_0)\cdot
\left(J-\frac{\kappa}{2}J^\mr{adv}\right)+e\tG^\mr{ret}_0\cdot\left(A-\frac{\kappa}{2}A^\mr{adv}\right),\nn
D^\mr{ret}_0\cdot\left(\jb+\frac{\kappa}{2}j\right)&=&eD^\mr{ret}_0\cdot\left(J-\frac{\kappa}{2}J^\mr{adv}\right)
-A+\frac{\kappa}{2}A^\mr{adv}.
\eea

According to the remark made at the end of Appendix \ref{raton} the effective action
$\Gamma_\mr{FBC}[J,J^\mr{adv},A,A^\mr{adv}]$ is obtained from $\Gamma_\mr{FBC}[J,J^\mr{adv},A,A^\mr{adv}]$
by the replacement of all retarded or advanced propagators by their near field version
as long as pair creation is neglected.

\subsection{Effective actions for two fields}\label{eftf}
The effective action which involves physical fields results from the Legendre
transformation in the variables $a$ and $j$ only. The generating functional
$\Re W[a,\ab,j,\jb]$ based on the two-point function \eq{inwobc} or \eq{inwfbc} is
\be
\Re W[a,\ab,j,\jb]=-\frac{\kappa}{2}(a,j)\cdot\begin{pmatrix}\tG^\mr{near}&e\tG^\mr{near}\cdot D_0^\mr{near}\cr
e\tG^\mr{near}\cdot D_0^\mr{near}&D_0\end{pmatrix}\cdot \begin{pmatrix}a\cr j\end{pmatrix}
+(a,j)\cdot\begin{pmatrix}W_a\cr W_j\end{pmatrix}
\ee
where the only boundary condition dependent pieces are 
\be\label{wtfa}
W_{\mr{OBC}a}=-\tG^\mr{ret}\cdot(\ab+eD_0^\mr{ret}\cdot\jb)~~~~~~W_{\mr{OBC}j}=-D^\mr{ret}\cdot\jb-eD_0^\mr{ret}\cdot\tG^\mr{ret}\cdot\ab,
\ee
or
\be\label{wtfj}
W_{\mr{FBC}a}=-\tG^\mr{near}\cdot(\ab+eD_0^\mr{near}\cdot\jb)~~~~~~W_{\mr{FBC}j}=-D_0\cdot\jb-eD_0^\mr{near}\cdot\tG^\mr{near}\cdot\ab.
\ee
The application of the inverse \eq{invbm} produces the effective action 
\be
\Gamma_\mr{FBC}[J,A]=-\frac{1}{2\kappa}(J,A)\cdot K^{-1}\cdot\begin{pmatrix}J\cr A\end{pmatrix}
+\frac{1}{\kappa}(J,A)\cdot K^{-1}\cdot\begin{pmatrix}W_a\cr W_j\end{pmatrix}
\ee
with
\be
K^{-1}=-\frac{1}{\kappa}\begin{pmatrix}\tG^{\mr{near}-1}_0+\tG^{\mr{near}-1}_0\cdot W^{(4)n}_D\cdot\tG^{\mr{near}-1}_0&-e\cr-e&D^{\mr{near}-1}_0\end{pmatrix}.
\ee

\section{Functional Renormalization Group in the CTP Formalism}\label{frgctp}
In the original version of the renormalization group method one follows
the evolution of coupling constants one-by-one in the framework of the
perturbation expansion \cite{grw} and exact results might be achieved by
resumming the perturbation expansion. The functional realization of the
renormalization group \cite{wetteq,morriseq} is based on one-loop evolution
equation and the exact results should arise by letting infinitely many term mixing
in the action. In this Appendix we outline briefly the evolution equation for the
effective action in the latter formalism which can easily be converted
into a numerical, non-perturbative algorithm to solve models.

Let us introduce a cutoff $k$ in the theory for the field variable $\phi$ 
in the quadratic part of the action, $S[\phi]\to S[\phi]+i\phi\cdot K_k\cdot\phi/2$.
The properties of the real operator $K$ are: (i) It should suppress all 
fluctuations for $k\to\infty$, ie. the eigenvalues of $K_\infty$ should be $\infty$.
(ii) The physical theory should be recovered for $k=0$ which is guaranteed by 
the condition $K_0=0$. (iii) For finite, non-vanishing $k$ $K_k$ should suppress 
the modes with momentum below $k$ ie. the eigenvalues of the translation invariant 
$K_k$ should be large for momenta $p<k$. The generating functional for the connected Green 
functions of the field is 
\be
e^{iW[j^+,j^-]}=\int D[\hhi]e^{iS[\hhi]-\hf\hhi\cdot\hat K\cdot\hhi+i\hj\cdot\hhi},
\ee
$\hhi=(\phi^+,\phi^-)$ and $S[\hhi]=S[\phi^+]-S^*[\phi^-]$.
The generating functional is equipped with UV regulator which is omitted for 
simplicity. The quadratic term in the exponent controls the fluctuations and
\be
\hat K=\begin{pmatrix}K&0\cr0&K\end{pmatrix}.
\ee
The evolution of the generating functional
\be
\dot W=\hf\Tr\left[\left(\fdd{W}{j^+}{j^+}+\fdd{W}{j^-}{j^-}+i\fd{W}{j^+}\fd{W}{j^+}
+i\fd{W}{j^-}\fd{W}{j^-}\right)\cdot\dot K\right]
\ee
is obtained by bringing the derivation with respect to $k$, denoted by a dot, into the 
functional integral. The parameterization \eq{jpar} of the external sources gives
\be\label{evolw}
\dot W=\Tr\left[\left(\fdd{W}{j}{j}-\kappa\fdd{W}{j}{\jb}+\frac{1+\kappa^2}{4}\fdd{W}{\jb}{\jb}
+\frac{i}{2}(\phi^+\phi^++\phi^-\phi^-)\right)\cdot\dot K\right].
\ee

We need the evolution of the effective action
\be
\Gamma[\phi,\phi^\mr{im}]=\Gamma[\phi,\phi^\mr{adv}]+i\Im\Gamma[\phi^\mr{im},\phi^\mr{adv~im}]
\ee
where we separate the real and imaginary parts
\bea
\Gamma[\phi,\phi^\mr{adv}]&=&\Re W[j,\jb]-\jb\cdot\phi^\mr{adv}-j\cdot\phi,\nn
\Im\Gamma[\phi^\mr{im},\phi^\mr{adv~im}]&=&\Im W[j,\jb]-\jb\cdot\phi^\mr{adv~im}-j\cdot\phi^\mr{im}
\eea
and introduce the variables
\bea
\phi+i\phi^\mr{im}&=&\fd{W[j,\jb]}{j},\nn
\phi^\mr{adv}+i\phi^\mr{adv~im}&=&\fd{W[j,\jb]}{\jb}.
\eea
The sources are real and therefore, two field variables are independent only.
We shall use $\phi$, $\phi^\mr{adv}$ as independent variables and introduce the functional
$\Gamma^\mr{im}[\phi,\phi^\mr{adv}]=\Im\Gamma[\phi^\mr{im},\phi^\mr{adv~im}]$ for the imaginary part.
The derivation of the evolution equation for the effective actions is straightforward
except that we have to keep track of the evolution of the relations among the 
independent and the dependent field variables. The expressions which follow 
will appear simpler by the help of the notations 
$\hhi=(\phi,\phi^\mr{adv})$, $\hhi^\mr{im}=(\phi^\mr{im},\phi^\mr{adv~im})$ adopted in the rest of this Appendix.

The equation \eq{wgmder} allows us to rewrite the evolution equation \eq{evolw} 
in terms of the effective action,
\bea\label{evolwe}
\dot\Gamma&=&\Tr\Biggl\{\Biggl[\left(\fdd{\Gamma}{\hhi}{\hhi}\right)^{-1}_{\phi\phi}
-\kappa\left(\fdd{\Gamma}{\hhi}{\hhi}\right)^{-1}_{\phi\phi^\mr{adv}}
+\frac{1+\kappa^2}{4}\left(\fdd{\Gamma}{\hhi}{\hhi}\right)^{-1}_{\phi^\mr{adv}\phi^\mr{adv}}\\
&&+i\left(\fdd{\Im\Gamma}{\hhi^\mr{im}}{\hhi^\mr{im}}\right)^{-1}_{\phi^\mr{im}\phi^\mr{im}}
-\kappa i\left(\fdd{\Im\Gamma}{\hhi^\mr{im}}{\hhi^\mr{im}}\right)^{-1}_{\phi^\mr{im}\phi^\mr{adv~im}}
+i\frac{1+\kappa^2}{4}\left(\fdd{\Im\Gamma}{\hhi}{\hhi}\right)^{-1}_{\phi^\mr{adv~im}\phi^\mr{adv~im}}
+\frac{i}{2}(\phi^+\phi^++\phi^-\phi^-)\Biggr]\cdot\dot K\Biggr\}.\nonumber
\eea
were the subscript indicates block-matrix elements of the second functional
derivative. Furthermore the disconnected contributions to the two-point functions
are given in terms of $\phi^\pm=\phi+i\phi^\mr{im}-(\kappa\mp1)(\phi^\mr{adv}+i\phi^\mr{adv~im})/2$.

The evolution equation \eq{evolwe} is not yet closed because it contains the dependent
field variables, too. To relate $\hhi^\mr{im}$ and $\hhi$ locally in the field configuration space
we introduce the derivative matrix
\be
\hat S_{ab}[\hhi]=\fd{\hhi_a}{\hhi^\mr{im}_b}.
\ee
The elimination of the variables $\hhi^\mr{im}$ is achieved by the help of the derivation
of the equations of motion
\be
-\hj=\fd{\Im\Gamma}{\hhi^\mr{im}}=\fd{\Gamma}{\hhi}
\ee
with respect to $\hhi^\mr{im}$,
\be
\fdd{\Im\Gamma}{\hhi^\mr{im}}{\hhi^\mr{im}}=\fdd{\Gamma}{\hhi}{\hhi}\cdot\hat S.
\ee
The derivation of this equation with respect to $\lambda$ yields the evolution equation for $\hat S$,
\be\label{evols}
\dot{\hat S}_{ab}=\left(\fdd{\Gamma}{\hhi}{\hhi}\right)^{-1}_{ac}
\left[\left(\fdd{\dot\Gamma}{\hhi}{\hhi}\right)_{cd}\hat S_{db}
-\hat S_{dc}\left(\fdd{\dot\Gamma^\mr{im}}{\hhi}{\hhi}\right)_{de}\hat S_{eb}
-\fd{\dot\Gamma^\mr{im}}{\hhi_d}\fd{\hat S_{dc}}{\hhi_e}(S^{-1})_{eb}\right].
\ee

After the solution of the problem posed by the dependent variable we return to the 
evolution equation \eq{evolwe} and we separate the tree-level contribution by 
the parameterization
\be
\Gamma=\tilde\Gamma+i\Im\tilde\Gamma+\frac{i}{2}(\phi^+\cdot\hat K\cdot\phi^++\phi^-\cdot\hat K\cdot\phi^-)
\ee
of the effective action. The real and imaginary parts of $\tilde\Gamma$ satisfy 
the evolution equations
\bea\label{evolgt}
\dot{\tilde\Gamma}&=&\Tr\left\{\left[\left(\fdd{\Re\tilde\Gamma}{\hhi}{\hhi}+A\right)^{-1}_{\phi\phi}
-\kappa\left(\fdd{\Re\tilde\Gamma}{\hhi}{\hhi}+A\right)^{-1}_{\phi\phi^\mr{adv}}
+\frac{1+\kappa^2}{4}\left(\fdd{\Re\tilde\Gamma}{\hhi}{\hhi}+A\right)^{-1}_{\phi^\mr{adv}\phi^\mr{adv}}
\right]\cdot\dot K\right\},\\
\dot{\tilde\Gamma}^\mr{im}&=&\Tr\left\{\left[\left(\fdd{\Re\Gamma}{\hhi}{\hhi}\cdot\hat S+B\right)^{-1}_{\phi^\mr{im}\phi^\mr{im}}
-\kappa\left(\fdd{\Re\Gamma}{\hhi}{\hhi}\cdot\hat S+B\right)^{-1}_{\phi^\mr{im}\phi^\mr{adv~im}}
+\frac{1+\kappa^2}{4}\left(\fdd{\Re\Gamma}{\hhi}{\hhi}\cdot\hat S+B\right)^{-1}_{\phi^\mr{adv~im}\phi^\mr{adv~im}}
\right]\cdot\dot K\right\},\nonumber
\eea
respectively, where
\bea
A_{ab}&=&\hf\left(\hat K_{ac}\fd{\hhi^\mr{im}_c}{\hhi_b}+\fd{\hhi^\mr{im}_c}{\hhi_a}\hat K_{cb}\right)
+\hhi_c\hat K_{cd}\fdd{\hhi^\mr{im}_d}{\hhi_a}{\hhi_b},\nn
B_{ab}&=&2\hat K_{ab}-2\fd{\hhi_c}{\hhi^\mr{im}_a}\hat K_{cd}\fd{\hhi_d}{\hhi^\mr{im}_b}
-\hhi_c\hat K_{cd}\fdd{\hhi_d}{\hhi^\mr{im}_a}{\hhi^\mr{im}_b}
-\fdd{\hhi_c}{\hhi^\mr{im}_a}{\hhi^\mr{im}_b}\hat K_{cd}\hhi_d.
\eea

The effective action of the physical system is obtained by integrating the 
system of equations \eq{evols}, \eq{evolgt} from $k\approx\infty$ where 
a perturbative initial condition is imposed down to $k=0$. 
The renormalized trajectory displays
the scale dependence of the effective action. In the case of QED one should
start with properly regulated theory eg. the effective action should be calculated
as in Eq. \eq{wqed} after the replacement $\hD^{-1}_0\to\hD^{-1}_0+T\hat K_\lambda$
in the action for the initial value of $\lambda$ but keeping the original,
$\lambda$-independent counterterms.
The projection $T$ into the transverse photon states in the suppression term 
renders the turning on of the interactions of photons with the fermion loops 
according to their momentum in a gauge invariant manner.

\end{document}